\documentclass{aa}
\usepackage{amsmath,amssymb,epsf}

\makeatletter
\def\thebibliography#1{\section*{REFERENCES}
 \addcontentsline{toc}{section}{REFERENCES}
 \list{}{\labelwidth\z@
         \leftmargin 1.5em
	 \itemsep \z@
	 \itemindent-\leftmargin}
 \small\raggedright
 \parindent\z@
 \parskip\z@ plus .1pt\relax
 \def\newblock{\hskip .11em plus .33em minus .07em}
 \sloppy\clubpenalty4000\widowpenalty4000
 \sfcode`\.=1000\relax
}

\def\@biblabel#1{}
\def\@bcite#1#2{(#1\if@tempswa , #2\fi)}
\def\@pcite#1#2{#1\if@tempswa , #2\fi}
\def\@citefmta#1#2{#1 (#2)}
\def\@citefmtb#1#2{#1 #2}
\let\citefmt=\@citefmta

\def\@citex[#1]#2{\if@filesw\immediate\write\@auxout{\string\citation{#2}}\fi
  \def\@citea{}\@cite{\@for\@citeb:=#2\do
    {\@citea\def\@citea{;\penalty\@m\ }\@ifundefined
    {b@\@citeb}{{\bf ?}\@warning
{Citation `\@citeb' on page \thepage \space undefined}}%
{\csname b@\@citeb\endcsname}}}{#1}}

\def\cite{\@ifnextchar [{\let\citefmt=\@citefmtb
                          \let\@cite=\@bcite\@tempswatrue \@citex}
                        {\let\citefmt=\@citefmtb
                          \let\@cite=\@bcite\@tempswafalse \@citex[]}}
\def\pcite{\@ifnextchar [{\let\citefmt=\@citefmtb
                          \let\@cite=\@pcite\@tempswatrue\@citex}
                        {\let\citefmt=\@citefmtb
                          \let\@cite=\@pcite\@tempswafalse\@citex[]}}
\def\scite{\@ifnextchar [{\let\citefmt=\@citefmta
                          \let\@cite=\@pcite\@tempswatrue\@citex}
                        {\let\citefmt=\@citefmta
                          \let\@cite=\@pcite\@tempswafalse\@citex[]}}
\makeatother

\def\hMpc{\ifmmode{h^{-1}{\rm Mpc}}\else{$h^{-1}$Mpc}\fi}

\def\d{{\rm d}}
\def\e{{\rm e}}
\def\bk{{\mathbf{k}}}

\def\bu{{\mathbf{u}}}
\def\bx{{\mathbf{x}}}
\def\by{{\mathbf{y}}}
\def\bz{{\mathbf{z}}}

\def\BE{{\mathbb{E}}}
\def\BR{{\mathbb{R}}}
\def\BV{{\mathbb{V}}}

\def\CB{{\cal B}}
\def\CC{{\cal C}}
\def\CD{{\cal D}}
\def\CX{{\cal X}}

\def\ii{{\'\i}}

\newcommand{\Beins}{\mbox{1\hspace*{-0.085cm}l}}

\newcommand{\aanda}{A\&A}

\newcommand{\apj}{ApJ}

\newcommand{\apjs}{ApJS}

\newcommand{\mnras}{MNRAS}
\newcommand{\nat}{Nat}

\begin{document}

\thesaurus{02(12.12.1, 12.03.4)}
\title{The geometry of second--order statistics -- biases in common estimators}
\author{Martin Kerscher}
\institute{Sektion Physik, Ludwig--Maximilians--Universit\"{a}t, \\ 
Theresienstra{\ss}e 37, D--80333 M\"{u}nchen, Germany\\
email: kerscher@stat.physik.uni-muenchen.de}

\date{Received 28 July 1998, accepted 26 October 1998}

\maketitle

\begin{abstract}
Second--order measures,  such as the  two--point correlation function,
are geometrical  quantities describing the clustering  properties of a
point distribution.   In this  article well--known estimators  for the
correlation integral  are reviewed  and their relation  to geometrical
estimators  for the  two--point correlation  function is  put forward.
Simulations illustrate the range of applicability of these estimators.
The  interpretation  of the  two--point  correlation  function as  the
excess of  clustering with respect  to Poisson distributed  points has
led to biases in  common estimators.  Comparing with the approximately
unbiased  geometrical  estimators,  we   show  how  biases  enter  the
estimators      introduced     by     {}\scite{davis:survey-twopoint},
{}\scite{landy:bias},   and   {}\scite{hamilton:towards}.    We   give
recommendations  for  the  application  of the  estimators,  including
details  of  the  numerical  implementation.  The  properties  of  the
estimators  of   the  correlation  integral  are   illustrated  in  an
application to  a sample of IRAS  galaxies.  It is found  that, due to
the limitations of  current galaxy catalogues in number  and depth, no
reliable determination of the  correlation integral on large scales is
possible.   In  the  sample   of  IRAS  galaxies  considered,  several
estimators  using different  finite--size corrections  yield different
results on  scales\footnote{Throughout this article  we measure length
in units  of \hMpc,  with $H_0  = 100h\ {\rm  km}\ {\rm  s}^{-1}\ {\rm
Mpc}^{-1}$.}  larger than 20\hMpc, while  all of them agree on smaller
scales.
\end{abstract}

\section{Introduction}

Second--order measures, also called two--point measures, are still one
of  the  major  tools  to  characterize the  spatial  distribution  of
galaxies  and clusters.  Probably  the best  known are  the two--point
correlation function $g(r)$  and the normed cumulant $\xi_2(r)=g(r)-1$
(e.g.~\pcite{peebles:lss}).  With  the mean number  density denoted by
$\overline{\rho}$,
\begin{equation}
\overline{\rho}^2 g(r)\ \d V(\bx_1) \d V(\bx_2)
\end{equation}
describes the  probability to find a  point in the  volume element $\d
V(\bx_1)$ {\em  and} another point  in $\d V(\bx_2)$, at  the distance
$r=\|\bx_1-\bx_2\|$, $\|\cdot\|$ is the Euclidean norm of a vector.
The              correlation              integral              $C(r)$
(e.g.~{}\pcite{grassberger:dimensions})  is   the  average  number  of
points  inside  a  ball of  radius  $r$  centred  on  a point  of  the
distribution; hence,
\begin{equation}
C(r) = \int_0^r\d s\ \overline{\rho}\ 4\pi s^2 g(s) .
\end{equation}
In Appendix~A we discuss other common two--point measures.

The   correlation   integral $C(r)$  and    the two--point correlation
function $g(r)$ are {\em defined} as ensemble averages.  If we want to
{\em estimate} $C(r)$  from one given point  set,  as provided by  the
spatial coordinates of galaxies, we have to  use volume averages which
yield an {\em estimator} $\widehat{C}(r)$.

Since all astronomical catalogues  are  spatially limited, i.e.\   the
observed  galaxies lie inside a  spatial domain $\CD$, we must correct
for boundary effects.
Estimators   of the    two--point    correlation  function   including
finite--size    corrections      have   been        proposed        by
{}\scite{hewett:estimation},          {}\scite{davis:survey-twopoint},
{}\scite{rivolo:two-point},                      {}\scite{landy:bias},
{}\scite{hamilton:towards},         {}\scite{szapudi:new},         and
{}\scite{ponsborderia:comparing}, to name only a few.
In a  recent   paper {}\scite{stoyan:improving}  introduced   improved
estimators  of point process  statistics, with special emphasis on the
accurate estimation of the density $\overline{\rho}$.

An  estimator  $\widehat{C}(r)$   is  called   {\em  unbiased} if  the
expectation of $\widehat{C}(r)$ equals the true value of $C(r)$:
\begin{equation}
\BE\  [ \widehat{C}(r) ] = C(r).
\end{equation} 
$\BE$ denotes the expectation value, the  average over realizations of
the   point   process\footnote{We assume  that  the   point process is
stationary.}.
An estimator is called  {\em consistent}\footnote{For an ergodic point
process an   unbiased   estimator is   also consistent.},  if   the
estimates  $\widehat{C}(r)$ obtained inside  a finite  sample geometry
$\CD$  from {\em one} space  filling realization, converge towards the
true value of $C(r)$, as the sample volume $|\CD|$ increases:
\begin{equation}
\widehat{C}(r)  \overset{|\CD|\rightarrow\infty}{\longrightarrow} C(r).
\end{equation}
We call  an estimator {\em ratio--unbiased}  if it is  the quotient of
two  unbiased quantities.  Whether  such a  quotient gives  a reliable
estimate  must  be  tested  .    Often  this  is  only  possible  with
simulations     (see    Sect.~\ref{sect:C-comparison},     see    also
{}\pcite{hui:biased}).

For the comparison of a simulated  point distribution with an observed
galaxy distribution within the same sample  geometry and with the same
selection   effects,  unbiasedness (or  consistency)   is  not a major
concern.  It is more  important that the  variance of the estimator is
small.  This  may give tighter  bounds on  the cosmological parameters
entering the simulations.

This       article    is       organized      as     follows.       In
Sect.~\ref{sect:corrint-estimators} we will review several  estimators
for  the correlation integral.    With simulations of  two drastically
different point process  models, namely a featureless  Poisson process
and  a highly structured  line segment  process,  the variance and the
bias of the estimators are investigated.
Closely connected to these estimators for the correlation integral are
the  geometrical  estimators for  the  two--point correlation function
which will be discussed in Sect.~\ref{sect:geometrical-twopoint-corr}.
Some popular   pair--count estimators  for the  two-point  correlation
function are considered in Sect.~\ref{sect:pair-estimators}. We derive
the geometrical properties of the pair--counts.  By comparing with the
geometrical estimators of   Sect.~\ref{sect:geometrical-twopoint-corr}
and with numerical examples we show how biases enter.
We comment on the improved estimators of {}\scite{stoyan:improving} in
Sect.~\ref{sect:improved}.
As  an   application,  we investigate   the  clustering  properties of
galaxies in  a volume  limited  sample of   the IRAS  1.2~Jy  redshift
catalogue in Sect.~\ref{sect:corrint-1.2jy}.
We  conclude  and  give  recommendation  for the  application  of  the
estimators in Sect.~\ref{sect:conclusion}.
In the Appendices we  summarize currently used two--point measures and
discuss some  details concerning  the numerical implementation  of the
estimators.

\section{Estimators for the correlation integral $C(r)$}
\label{sect:corrint-estimators}

Consider a set of points $\CX=\{\bx_i\}_{i=1}^{N}$, $\bx_{i}\in\BR^3$,
supplied by the  redshift coordinates of a  galaxy  or cluster survey.
All points $\bx_{i}$ are inside the sample geometry $\CD$.

\subsection{The na{\"\i}ve estimator for $C(r)$}

\begin{figure}
\begin{center}
\epsfxsize=8cm
\begin{minipage}{\epsfxsize} \epsffile{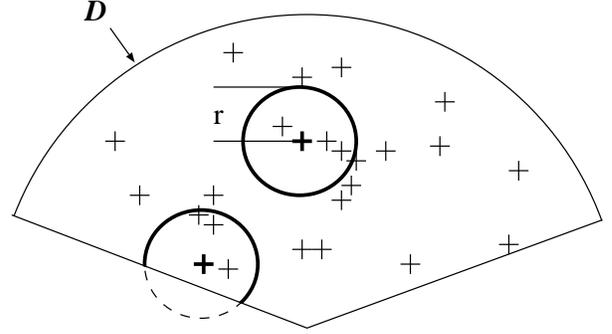} 
\end{minipage} 
\end{center}
\caption{\label{fig:biased-C}       In   the  na{\"\i}ve     estimator
$\widehat{C_0}(r)$  all   points    are used    as  centres  for   the
determination of $N_i(r)$. $N_i(r)$ is  underestimated for points near
the boundary of $\CD$.}
\end{figure}
The na{\"\i}ve and   biased estimator   of the  correlation   integral
$\widehat{C_0}(r)$ is defined by
\begin{equation}
\widehat{C_0}(r) = \frac{1}{N} \sum_{i=1}^{N} N_i(r) ,
\end{equation}
where
\begin{equation}
\label{eq:local-count}
N_i(r) = \sum_{j=1; j\ne i}^{N} \Beins_{[0,r]}(\|\bx_i-\bx_j\|)
\end{equation}
is the number of points in a sphere with radius $r$ around the point 
$\bx_i$.
\begin{equation}
\Beins_{A}(x) = 
\begin{cases}
1 & \text{for } x \in A ,\\
0 & \text{for } x \notin  A 
\end{cases}
\end{equation}
denotes the indicator function of the set $A$.
$\widehat{C_0}(r)$  is the mean value  of $N_i(r)$,  averaged over all
points $\bx_i$.  For points $\bx_i$ near the boundary of $\CD$ and for
large  radii  $r$  in  particular  the number  of   points $N_i(r)$ is
underestimated,  and   $\widehat{C_0}(r)$  is biased  towards  smaller
values (see Fig.~\ref{fig:biased-C}).

\subsection{Minus--estimators for $C(r)$}
\label{sect:minus-C}

\begin{figure}
\begin{center}
\epsfxsize=8cm
\begin{minipage}{\epsfxsize} \epsffile{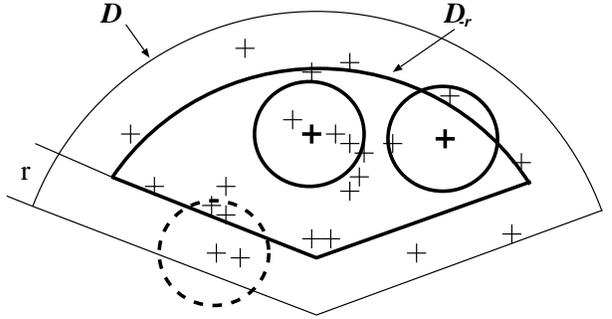} 
\end{minipage} 
\end{center}
\caption{\label{fig:minus-C}   In the  minus--estimators only   points
inside   $\CD_{-r}$ are taken  into  account  in  the determination of
$N_i(r)$.}
\end{figure}
As an obvious restriction, only points, further than $r$ away from the
boundary of $\CD$ are used as centres for the calculation of $N_i(r)$.
Doing so we make sure, that  we see all  data points inside the sphere
of radius $r$ around a point $\bx_{i}$.
$\CD_{-r}$ is the shrunken window (see Fig.~\ref{fig:minus-C})
\begin{equation}
\CD_{-r} = 
\{\by\in\CD~:~\CB_r(\by)\subset\CD\},
\end{equation}
where $\CB_r(\by)$ denotes  a sphere of radius $r$  centred on $\by$,
and
\begin{equation}
\label{eq:Nr-def}
N_r = \sum_{i=1}^{N} \Beins_{\CD_{-r}}(\bx_i) .
\end{equation}
yields the number of points inside $\CD_{-r}$
The minus--\-es\-ti\-ma\-tor $\widehat{C_1}(r)$ reads:
\begin{equation}
\widehat{C_1}(r) = \frac{1}{N_r}\ 
\sum_{i=1}^{N} \Beins_{\CD_{-r}}(\bx_i)\ N_i(r) \quad\text{ for } N_r>0.
\end{equation}
In  the case of   stationary    point  processes this estimator     is
ratio--un\-biased (e.g.\ {}\pcite{baddeley:3dpoint}).
However,  for large  radii  only a  small  fraction of  the points  is
included as  centres.  Therefore, we are  limited to scales  up to the
radius of  the largest sphere  that lies completely inside  the sample
geometry.  With this  estimator we do not have  to make any assumption
about the  distribution of points  outside the window $\CD$.   This is
important for the  investigation of inhomogeneous, scale--invariant or
``fractal'' point distributions.
Pietronero  and  coworkers  employed   this  type  of  estimator  (see
Appendix~A and {}\pcite{labini:scale}).

Let  us  introduce another  variant  of the  minus--\-es\-ti\-ma\-tor,
which  also does  not require  any assumption  about the  missing data
outside the sample window $\CD$.
An unbiased  estimator  of the  number  density is given by
\begin{equation}
\label{eq:stichprobendichte}
\widehat{\rho_1} = \frac{N}{|\CD|} ,
\end{equation}
and an alternative ratio--unbiased minus--estimator may be defined by
\begin{equation}
\widehat{C_2}(r) = \frac{1}{\widehat{\rho_1}\ |\CD_{-r}|} \ 
\sum_{i=1}^{N} \Beins_{\CD_{-r}}(\bx_i)\ N_i(r) 
\quad\text{ for } |\CD_{-r}|>0.
\end{equation} 
$\widehat{C_2}(r)$ differs from $\widehat{C_1}(r)$ in that we estimate
the    number    density     with    $N_r/|\CD_{-r}|$    instead    of
$\widehat{\rho_1}$, and an estimate of $\overline{\rho}$ from a larger
volume than in  $\widehat{C_1}(r)$ is used. This may  be important, if
the  galaxy  catalogue  under  consideration  is centred  on  a  large
cluster.   Then   $N_r\ge|\CD_{-r}|~\widehat{\rho_1}$,  and  therefore
$\widehat{C_1}(r)$   systematically  underestimates   the  correlation
integral $C(r)$.
On the other hand, in  $\widehat{C_1}(r)$ the same points are used for
the determination of the  numerator and denominator, which empirically
yields  a reduced variance.   In Sect.~\ref{sect:comparisonC}  we will
see that the large variance of $\widehat{C_2}(r)$ makes this estimator
rather useless.

\subsection{Ripley--estimator for $C(r)$}
\label{sect:ripley}

\begin{figure}
\begin{center}
\epsfxsize=8cm
\begin{minipage}{\epsfxsize} \epsffile{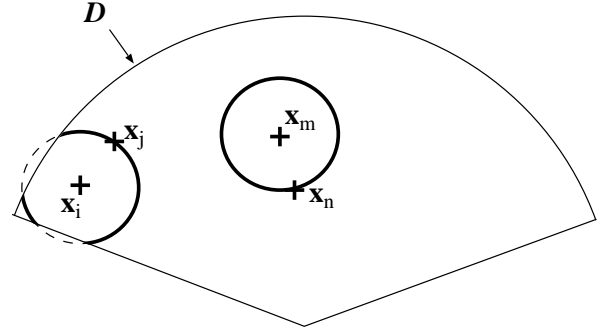} 
\end{minipage} 
\end{center}
\caption{\label{fig:local-weight} The local weight $\omega_l(\bx_m,s)$
equals unity  for the point  $\bx_m$ with $s=\|\bx_m-\bx_n\|$.  At the
point $\bx_i$ with $s=\|\bx_i-\bx_j\|$ the local weight is larger than
unity.}
\end{figure}

The   Ripley--estimator {}\cite{ripley:second-order}  uses  all points
inside $\CD$   as     centres   for    the   counts  $N_i(r)$     (see
Eq.~\ref{eq:local-count}).    The   bias   in $\widehat{C_0}(r)$    is
corrected with weights:
\begin{multline}
\label{eq:ripley}
\widehat{C_3}(r) = \frac{1}{N} \sum_{i=1}^{N}  \sum_{j=1; j\ne i}^{N} 
\Beins_{[0,r]}(\|\bx_i-\bx_j\|) \times \\
\times\ \omega_l(\bx_i,\|\bx_i-\bx_j\|)\ \omega_g(\|\bx_i-\bx_j\|) ,
\end{multline}
with the {\em local} pair weight {}\cite{ripley:second-order}
\begin{equation}
\label{eq:localweight}
\omega_l(\bx_i,s) =
\begin{cases}
\frac{4\pi s^2}{\text{area}(\partial\CB_{s}(\bx_i)\cap\CD)} 
& \text{for } \partial\CB_{s}(\bx_i)\cap\CD \ne \emptyset, \\
0 & \text{for } \partial\CB_{s}(\bx_i)\cap\CD = \emptyset,
\end{cases}
\end{equation}
inversely  proportional to  the  part of  the  spherical surface  with
radius  $s$  around the  point  $\bx_1$  which  is inside  the  survey
boundaries (see  Fig~\ref{fig:local-weight}).  $\partial\CB_s(\bx)$ is
the surface  of the sphere  $\CB_s(\bx_i)$ with radius $s$  centred on
$\bx_i$.
With $\omega_l(\bx_i,s)$ we correct  {\em locally} for possible points
at distance $s$ outside the sample geometry $\CD$.

\begin{figure}
\begin{center}
\epsfxsize=4cm
\begin{minipage}{\epsfxsize} \epsffile{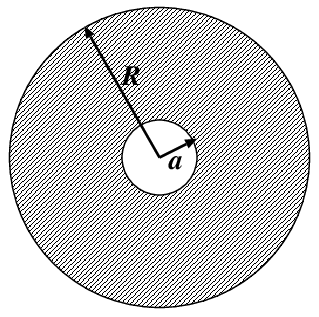} 
\end{minipage} 
\end{center}
\caption{\label{fig:global-weight}    The shaded  area  marks  the set
$\{\bx\in\CD~|~\partial\CB_s(\bx)\cap\CD\ne\emptyset\}$     for      a
spherical sample geometry $\CD=\CB_R$, with $s=R+a$.}
\end{figure}

The {\em global} weight
\begin{equation}
\omega_g(s) =
\frac{|\CD|}{|\{\bz\in\CD~|~\partial\CB_s(\bz)\cap\CD\ne\emptyset\}|} ,
\end{equation}
was   introduced  by  {}\scite{ohser:estimators}.    $\omega_g(s)$  is
inversely proportional  to the  volume occupied by  points $\bz\in\CD$
for  which  the  surface  $\partial\CB_s(\bz)$ intersects  the  sample
geometry $\CD$ (see Fig.~\ref{fig:global-weight}).
In typical sample geometries  the {\em global} weight $\omega_g(s)$ is
equal  to  unity   up  to  fairly  large  radii   $s$.   For  example,
$\omega_g(s)$  exceeds unity  only  for $s>R$  in  a spherical  sample
geometry  with  radius  $R$ (see  Fig.~\ref{fig:global-weight}).   For
$r<\max\{s\in\BR^+              \text{              with             }
|\{\bx\in\CD~|~\partial\CB_s(\bx)\cap\CD\ne\emptyset\}|>0\}$,
$\widehat{C_3}(r)$ is ratio--unbiased {}\cite{ohser:estimators}.

\subsection{Ohser and Stoyan estimators for $C(r)$}
\label{sect:ohser-stoyan-C}

\begin{figure}
\begin{center}
\epsfxsize=8cm
\begin{minipage}{\epsfxsize} \epsffile{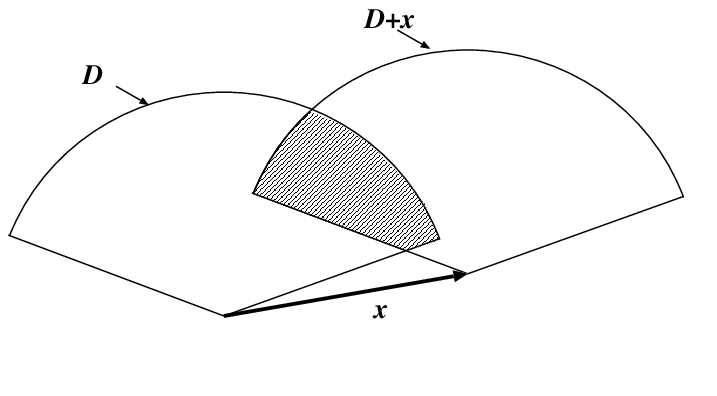} 
\end{minipage} 
\end{center}
\caption{\label{fig:setcovariance} 
The set--covariance $\gamma_\CD(\bx)$ is the volume of the shaded set 
$\CD\cap\CD+\bx$.}
\end{figure}
Another  estimator using   a weighting strategy  of    point pairs was
proposed by {}\scite{ohser:onsecond}:
\begin{multline}
\widehat{C_4}(r) = \frac{1}{N} \sum_{i=1}^{N} \sum_{j=1; j\ne i}^{N} 
\Beins_{[0,r]}(\|\bx_i-\bx_j\|)\ \frac{|\CD|}{\gamma_\CD(\bx_i-\bx_j)} \\
\text{if } \gamma_\CD(\bx_i-\bx_j) > 0 
\text{ for all } \|\bx_i-\bx_j\|<r . 
\end{multline}
Here     the    pair--weight    is     equal    to     the    fraction
$|\CD|/\gamma_\CD(\bx)$,     with     the     set--covariance     (see
Fig.~\ref{fig:setcovariance})
\begin{equation}
\label{eq:setcov}
\gamma_\CD(\bx) = |\CD \cap \CD+\bx|.
\end{equation}
$\CD+\bx$  is the  sample geometry  shifted by  the vector  $\bx$, and
$\gamma_\CD(\bx)$ is  the volume of  the intersection of  the original
sample with the shifted sample.
This  estimator  is ratio--unbiased   for stationary  point processes;
isotropy is not needed in the proof {}\cite{ohser:onsecond}.

Closely related to the estimator $\widehat{C_4}(r)$ is its isotropized
counterpart {}\cite{ohser:onsecond}:
\begin{multline}
\widehat{C_5}(r) = \frac{1}{N} 
\sum_{i=1}^{N}  \sum_{j=1; j\ne i}^{N} \Beins_{[0,r]}(\|\bx_i-\bx_j\|)\ 
\frac{|\CD|}{\overline{\gamma_\CD}(\|\bx_i-\bx_j\|)} ,\\
\text{for } \overline{\gamma_\CD}(r) > 0 ,
\end{multline}
where $\overline{\gamma_\CD}(r)$ is the isotropized set--covariance:
\begin{equation}
\label{eq:isotrop-setcov}
\overline{\gamma_\CD}(r) = \frac{1}{4\pi}\int_0^{2\pi}\int_0^\pi\
 \sin(\theta)\d\theta \d\phi\ \gamma_\CD(\bx(r,\theta,\phi)).
\end{equation}

\subsection{Comparison of the estimators for $C(r)$}
\label{sect:comparisonC}

Since  the  estimators  for  $C(r)$  considered above  are  only  {\em
ratio--unbiased}, we  have tested  whether they give  reliable results
with two drastically different examples of point processes.  This also
enables  us  to compare  the  variances  of  the estimators.   Several
analytical  approaches  have  been  put  forward  to  investigate  the
variance of estimators for  two--point measures.  The majority of them
relies       on        Poisson       or       binomial       processes
(e.g.~{}\pcite{ripley:spatial}, and  {}\pcite{landy:bias}, see however
{}\pcite{stoyan:variance}, and {}\pcite{bernstein:variance}).
A similar  numerical comparison of estimators  for two--point measures
in     the     two--dimensional      case     was     performed     by
{}\scite{doguwa:edge-corrected}.

\label{sect:C-comparison}
\begin{figure*}
\begin{center}
\epsfxsize=16cm
\begin{minipage}{\epsfxsize} \epsffile{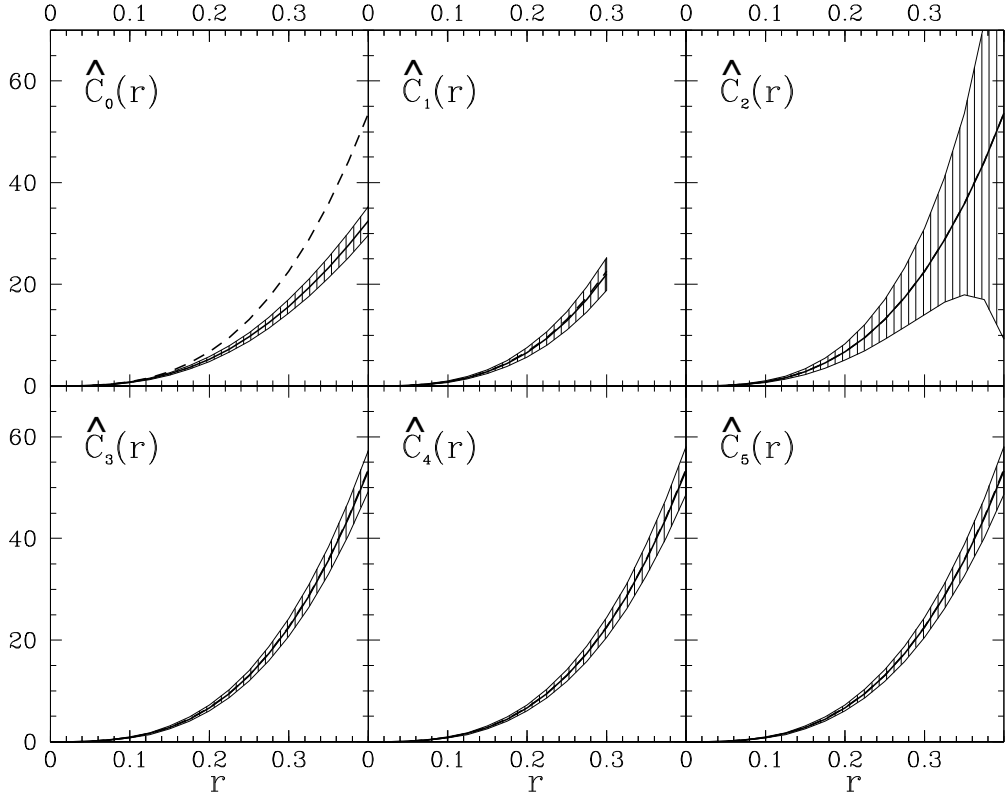} 
\end{minipage} 
\end{center}
\caption{\label{fig:poisson-Ccompare}  A comparison of  the estimators
$\widehat{C_0}(r)$  to $\widehat{C_5}(r)$ for  a Poisson  process with
number density $\overline{\rho}=200$ in  the unit box.  The solid line
marks  the  sample  mean,  the  shaded area  is  the  1$\sigma$--range
estimated from  10,000 realizations, and  the dashed line is  the true
$C_P(r)$.}
\end{figure*}
As a simple point process model  showing  no large--scale structure we
study the behaviour of the estimators for a  Poisson process with mean
number  density $\overline{\rho}$. The  mean value  of the correlation
integral is then
\begin{equation}
C_P(r) = \overline{\rho}\ \frac{4\pi}{3} r^3 .
\end{equation}
In Fig.~\ref{fig:poisson-Ccompare}   a  numerical   comparison of  the
estimators $\widehat{C_0}(r)$  to  $\widehat{C_5}(r)$  for a   Poisson
process   with $\overline{\rho}=200$ in the  unit  cube is shown.  The
mean and the  variance were determined from   10,000 realizations.  As
expected, a strong bias towards lower values  for large $r$ is seen in
$\widehat{C_0}(r)$;    the   other   estimators  do     not  show  any
bias.  $\widehat{C_1}(r)$   is     defined  only for     samples  with
$N_r>0$. Since there were samples with  $N_r=0$ for $r\ge0.325$ within
the 10,000 realizations of the  Poisson process, $\widehat{C_1}(r)$ is
shown only for radii smaller than $0.325$.

\begin{figure}
\begin{center}
\epsfxsize=8cm
\begin{minipage}{\epsfxsize} \epsffile{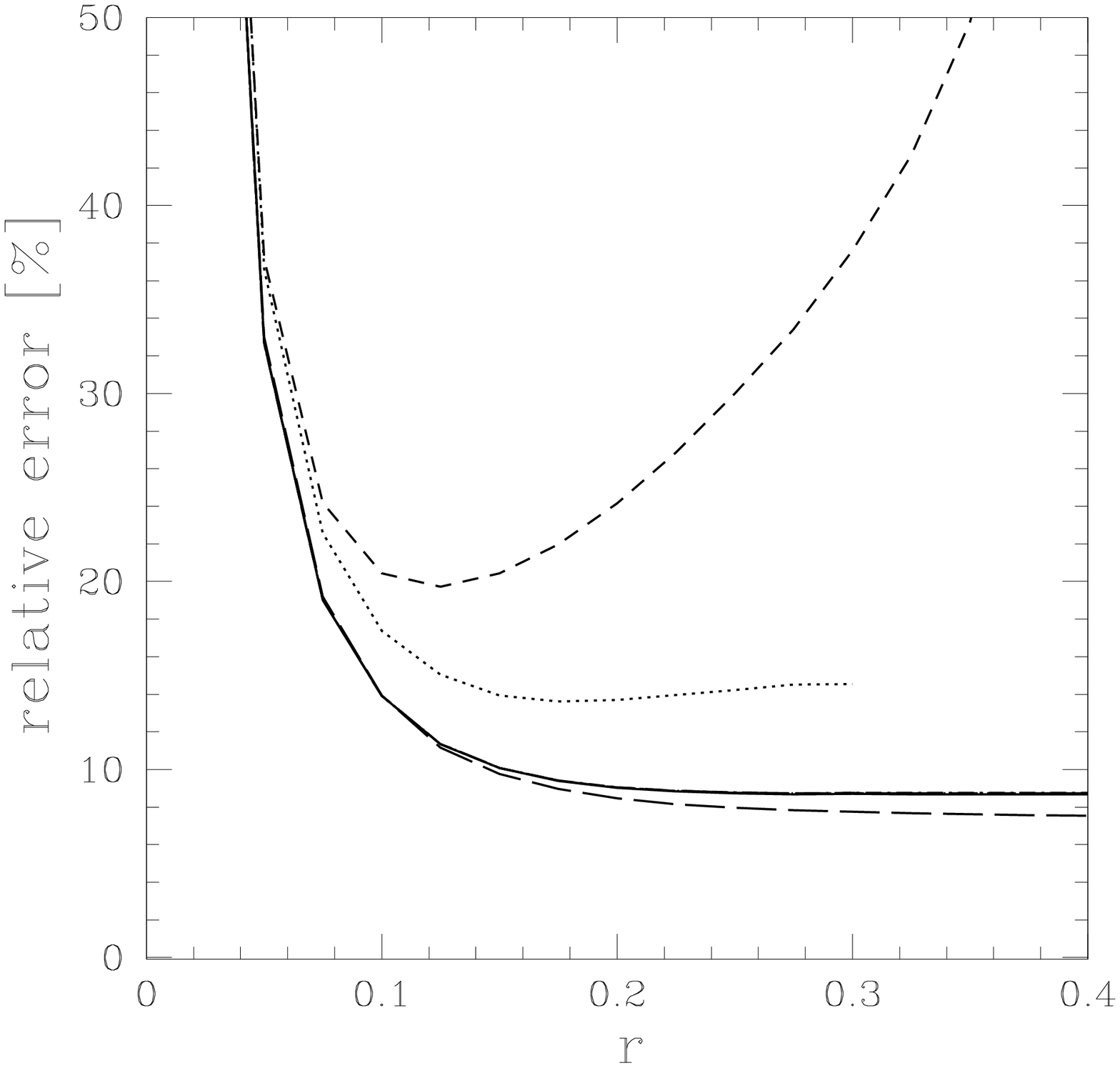}  
\end{minipage} 
\end{center}
\caption{\label{fig:poisson-errors}  The   comparison of  the relative
errors of  the estimators for  a  poisson process with  number density
$\overline{\rho}=200$  in   the unit box:  $\widehat{C_0}(r)$ (solid),
$\widehat{C_1}(r)$       (dotted),  $\widehat{C_2}(r)$ (short dashed),
$\widehat{C_3}(r)$     (long dashed),  $\widehat{C_4}(r)$       (short
dashed--dotted, on  top of    $\widehat{C_0}(r)$),  $\widehat{C_5}(r)$
(long dashed--dotted, on top of $\widehat{C_0}(r)$).}
\end{figure}

Looking at the absolute  errors in Fig.~\ref{fig:poisson-Ccompare}, we
see that the minus--estimators  exhibit larger errors than  the others
in particular,  $\widehat{C_2}(r)$  becomes  useless  on larger scales.
The   relative errors (the standard  error  per  mean value) exhibit a
``shot noise'' peak for small $r$ (see Fig.~\ref{fig:poisson-errors}).
All the estimators using weighting schemes show comparable errors, but
especially  for  large  $r$, the  Ripley  estimator $\widehat{C_3}(r)$
gives the smallest errors.

\begin{figure*}
\begin{center}
\epsfxsize=16cm
\begin{minipage}{\epsfxsize} \epsffile{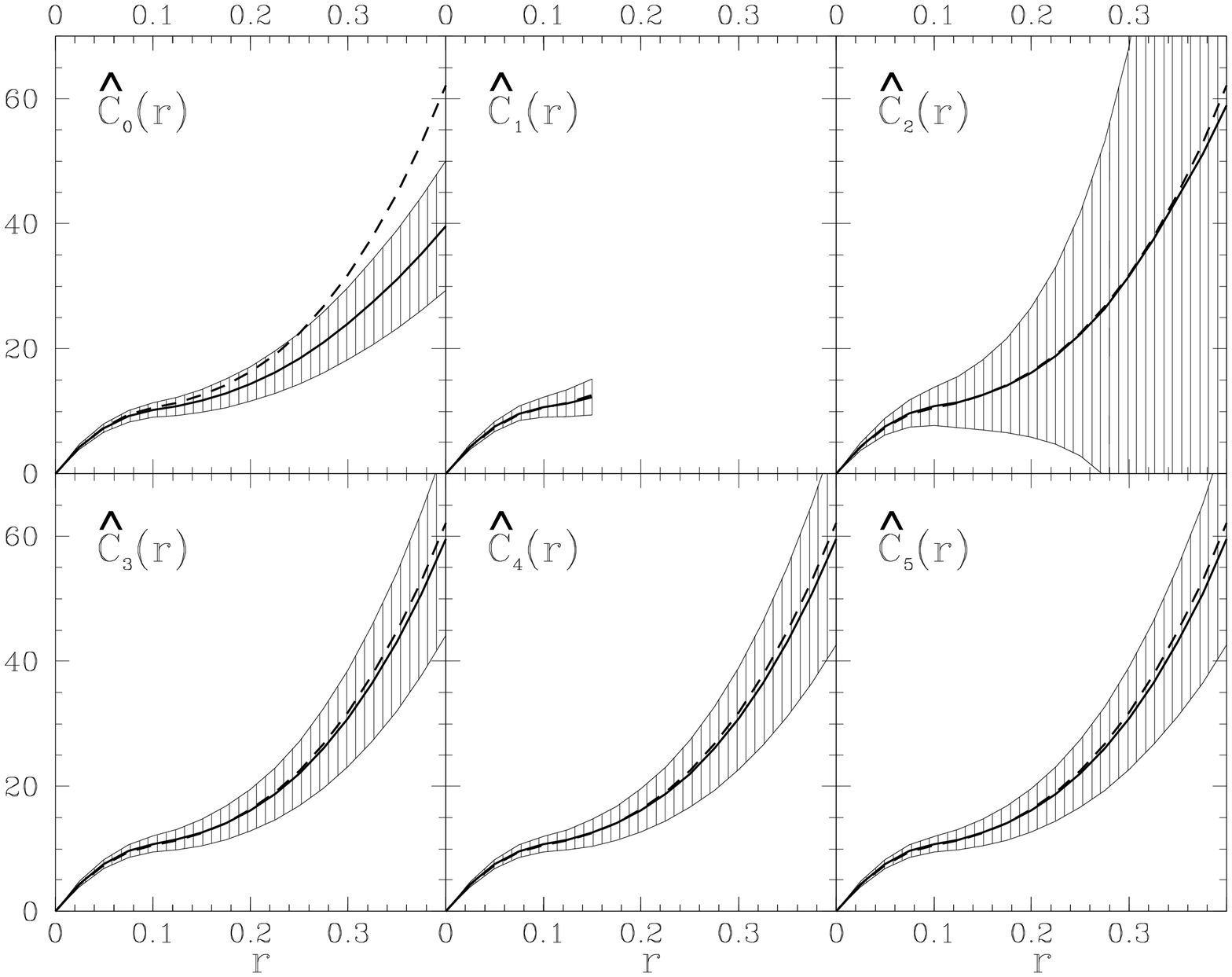} 
\end{minipage} 
\end{center}
\caption{\label{fig:line-Ccompare} A comparison of the estimators with
number density $\widehat{C_0}(r)$ to  $\widehat{C_5}(r)$ for  a random
distribution of points  with  $\overline{\rho}=200$ on  line  segments
with lenght $l=0.1$ and segment  density $\overline{\rho}_s=20$ in the
unit  box.  The solid  line marks the sample  mean, the shaded area is
the  1$\sigma$--range estimated  from   10,000 realizations,  and  the
dashed line is the true $C_S(r)$.}
\end{figure*}
To investigate the performance of the estimators for highly structured
and  clustered point   process  models,   we  study   points  randomly
distributed on   line    segments which  are     themselves  uniformly
distributed in  space and direction.  From {}\scite{stoyan:stochgeom},
p.~286 (see also {}\pcite{martinez:searching}) we obtain:
\begin{equation}
C_S(r) = 
\begin{cases}
\overline{\rho}\ \frac{4\pi}{3}r^3 + \frac{\overline{\rho}}{\overline{\rho}_s}
\left( \frac{2r}{l} - \frac{r^2}{l^2} \right) 
& \text{ for } r < l\\
\overline{\rho}\ \frac{4\pi}{3}r^3 + \frac{\overline{\rho}}{\overline{\rho}_s}
& \text{ for } r \ge l ;\\
\end{cases}
\end{equation}
$l$ is the length of  the line segments and $\overline{\rho}_s$ is the
mean   number   density   of  line   segments;   $l\overline{\rho}_s$,
$\overline{\rho}/(l\overline{\rho}_s)$,  $\overline{\rho}$  denote the
mean length density,  the mean number of points  per line segment, and
the mean  number density in  space, respectively. A similar  model for
the     distribution      of     galaxies     is      discussed     by
{}\scite{buryak:correlation}.
In Fig.~\ref{fig:line-Ccompare} we  compare the mean and  the variance
of the estimators  for 10,000 realizations  of a line segment  process
with $\overline{\rho}=200$, $l=0.1$ and $\overline{\rho}_s=20$.

\begin{figure}
\begin{center}
\epsfxsize=8cm
\begin{minipage}{\epsfxsize} \epsffile{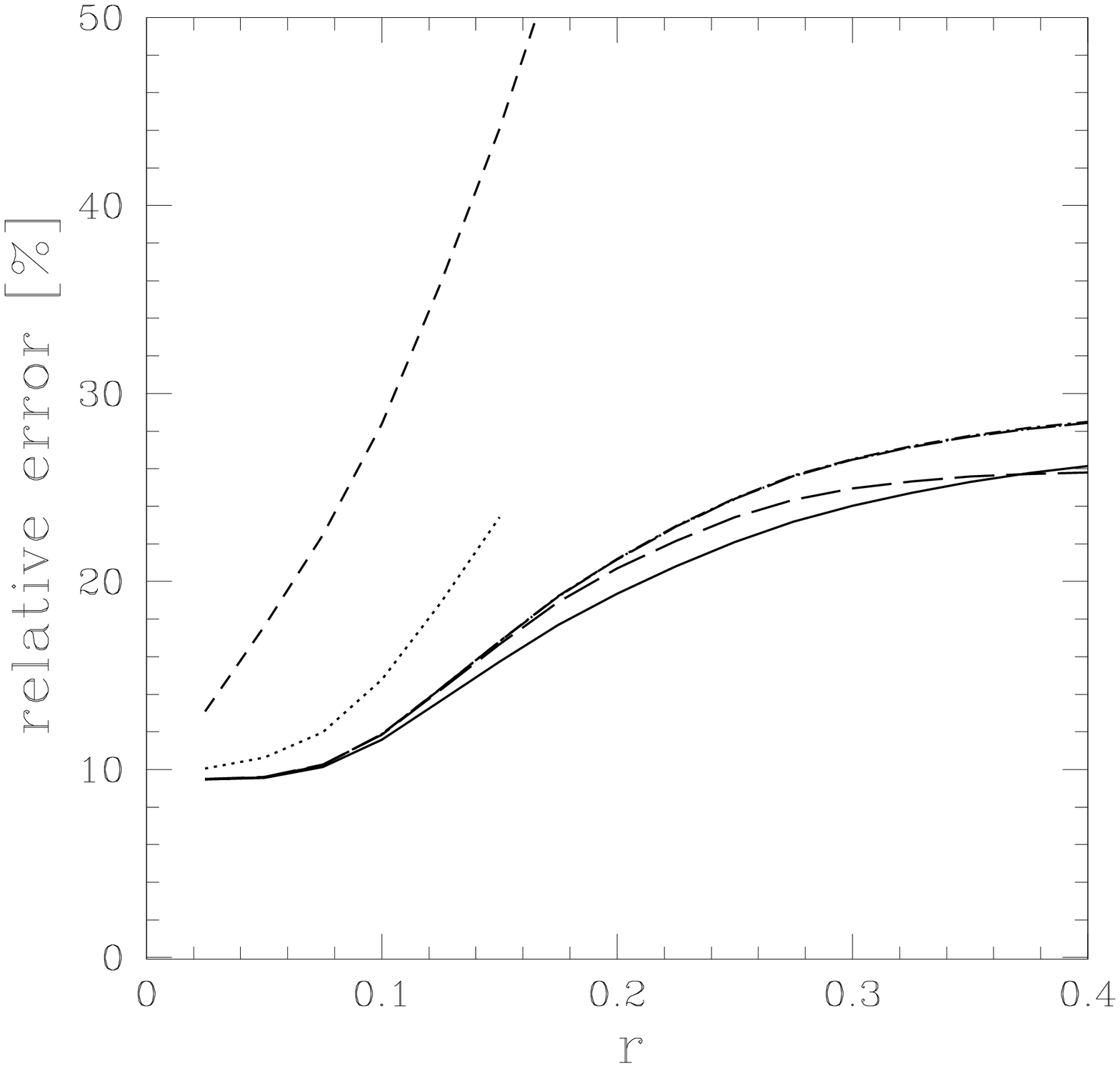} 
\end{minipage} 
\end{center}
\caption{\label{fig:line-errors} The comparison of the relative errors
of the estimators  for a  random distribution  of points  with  number
density $\overline{\rho}=200$ on line segments with lenght $l=0.1$ and
segment density       $\overline{\rho}_s=20$  in   the    unit    box:
$\widehat{C_0}(r)$     (solid),     $\widehat{C_1}(r)$       (dotted),
$\widehat{C_2}(r)$ (short dashed), $\widehat{C_3}(r)$ (long   dashed),
$\widehat{C_4}(r)$  (short dashed--dotted), $\widehat{C_5}(r)$   (long
dashed--dotted, on top of $\widehat{C_4}(r)$).}
\end{figure}

As before, $\widehat{C_0}(r)$ shows a strong bias on large scales, but
also the other estimators include  some bias towards smal\-ler values.
Some of the   random samples showed  $N_r=0$ for  $r<0.175$, therefore
$\widehat{C_1}(r)$ is given only for smaller radii.
Comparing             Fig.~\ref{fig:line-Ccompare}                 and
Fig.~\ref{fig:poisson-Ccompare} we   see that  this  clustered   point
distribution leads   to  a significantly  larger  variance  (see  also
{}\pcite{stoyan:inequalities}).
The relative errors (Fig.~\ref{fig:line-errors})  on large  scales are
nearly  twice as large as in  the case of a   Poisson process with the
same  number density  (Fig.~\ref{fig:poisson-errors}).  Since  we  are
looking  at  a clustered   distribution, the  ``shot noise''  peak  is
shifted   to    very     small       radii,    not  visible         in
Fig.~\ref{fig:line-errors}.     Again,       the      minus--estimator
$\widehat{C_2}(r)$ becomes  unreliable for large $r$.   The estimators
using weighting  schemes display a  significantly smaller  variance on
all scales, whereas  the Ripley estimator $\widehat{C_3}(r)$ gives the
smallest variance on large scales.
Simulations with different  parameters $l$ and $\overline{\rho_s}$ led
to the same conclusions.

A possible explanation why $\widehat{C_3}(r)$ shows a smaller variance
than      $\widehat{C_4}(r)$      and     $\widehat{C_5}(r)$      (see
Figs.~\ref{fig:poisson-errors} and  \ref{fig:line-errors}) is that the
local weight $\omega_l(\bx_i,s)$  used in $\widehat{C_3}(r)$ is larger
than unity only for a point  $\bx_i$ with another point at distance $s
(\le r)$ {\em  and} with $\bx_i$ closer to the  boundary of the sample
than  $s$.   The  weight  equals  unity for  all  other  point  pairs.
Contrary,    in   $\widehat{C_4}(r)$   and    $\widehat{C_5}(r)$   the
corresponding weights are larger than unity for all point pairs.  Each
of these  three estimators  is ratio--unbiased, hence,  correcting for
finite size effects,  but a frequent use of  weights larger than unity
increases  the   variance.   {}\scite{colombi:effects}  calculate  the
weights used in the estimation of the factorial moments minimizing the
variance of the factorial moments (see also {}\pcite{szapudi:cosmic}).

\section{Geometrical estimators for the two--point correlation 
function $g(r)$}
\label{sect:geometrical-twopoint-corr}

In contrast to estimators for the correlation integral, all estimators
of  the  two--point correlation  function  using  a  finite bin  width
$\Delta$ are biased.   A property similar to unbiasedness  is that the
expectation of such an estimator converges towards the true mean value
of $g(r)$  for $\Delta\rightarrow0$.  We call  this {\em approximately
unbiased}.

In this section we  discuss estimators for two--point correlation 
function $g(r)=1+\xi_2(r)$ which can
be derived from  the estimators for the correlation  integral given in
Sect.~\ref{sect:corrint-estimators}, by using the relation
\begin{equation}
\overline{\rho}\ 4\pi r^2\ g(r) = \overline{\rho}\ 4\pi r^2 (1+\xi_2(r)) = 
\frac{\d }{\d r}C(r).
\end{equation}

\subsection{The na{\"\i}ve estimator for $g(r)$}
\label{sect:naiv-g}

In  analogy  to   the  estimator  $\widehat{C_0}(r)$    we obtain  the
na{\"\i}ve estimator $\widehat{g_0}(r)$ for the two--point correlation
function $g(r)$:
\begin{equation}
\widehat{g_0}(r) = \frac{1}{N} \sum_{i=1}^N 
\frac{n_i^\Delta(r)}{4\pi r^2\Delta\ \widehat{\rho_1}}\ ,
\end{equation}
where
\begin{align}
\label{eq:def-nidelta}
n_i^\Delta(r) & =
\sum_{j=1, j\ne i}^N  \Beins_{[r,r+\Delta]}(\|\bx_i-\bx_j\|) \\
& = N_i(r+\Delta) - N_i(r) \ ,\nonumber
\end{align}
is the number  of points  in the  shell with  radius in $[r,r+\Delta]$
around  a point $\bx_i$.  $\widehat{\rho_1}=\tfrac{N}{|\CD|}$ provides
an estimate of the mean number density $\overline{\rho}$. The quotient
$\tfrac{n_i(r)}{\Delta}$ approximates $\tfrac{\d N_i(r)}{\d r}$:
\begin{equation}
4\pi r^2\widehat{\rho_1}\ \widehat{g_0}(r) \
\overset{\Delta\rightarrow0}{\longrightarrow} \
\frac{\d }{\d r}\widehat{C_0}(r)\ .
\end{equation}
Similar to   $\widehat{C_0}(r)$, $\widehat{g_0}(r)$ underestimates the
two--point correlation function $g(r)$.

\subsection{Minus--estimators for $g(r)$}
\label{sect:minus-g}

The minus--estimators for $g(r)$ are defined as follows:
\begin{align}
\widehat{g_1}(r) & = 
\frac{1}{N_r}\ \sum_{i=1}^N \Beins_{\CD_{-r}}(\bx_i)
\frac{n_i^\Delta(r)}{4\pi r^2\Delta\ \widehat{\rho_1}}\ , \\
\widehat{g_2}(r) & = 
\frac{1}{|\CD_{-r}|~\widehat{\rho_1}}\ \sum_{i=1}^N 
\Beins_{\CD_{-r}}(\bx_i)
\frac{n_i^\Delta(r)}{4\pi r^2\Delta\ \widehat{\rho_1}}\ ,
\end{align}
with $N_r>0$  and  $|\CD_{-r}|>0$.   As  in Sect.~\ref{sect:naiv-g} we
obtain   the  minus--estimators  for   $g(r)$  as derivatives   of the
minus--estimators for the correlation integral.
Therefore,   $\widehat{g_1}(r)$   and     $\widehat{g_2}(r)$       are
ratio--unbiased  in the  limit  $\Delta\rightarrow0$.   Pietronero and
coworkers  use $\widehat{\rho_1}\ {}\widehat{g_1}(r)$  to estimate the
conditional density $\Gamma(r)$.

\subsection{Rivolo estimator for $g(r)$}
\label{sect:rivolo}

{}\scite{rivolo:two-point}   suggested  a   pair--weighted  estimator,
defined as:
\begin{align} \label{eq:rivolo}
\widehat{g_3}(r) & = \frac{1}{N} \sum_{i=1}^N 
\frac{n_i^\Delta(r)}{4\pi r^2\Delta\ \widehat{\rho_1}}\ 
\omega_l(\bx_i, r) \\
& = \frac{|\CD|}{N^2} \sum_{i=1}^N 
\frac{n_i^\Delta(r) / 
\Delta}{\text{area}(\partial\CB_r(\bx_i)\cap\CD)} . \nonumber
\end{align}
For small $\Delta$ we obtain
\begin{multline}
\frac{n_i(r)^\Delta}{\Delta}\ \omega_l(\bx_i, r)\ 
\overset{\Delta\rightarrow0}{\longrightarrow} \\
\sum_{j=1; j\ne i}^{N} \delta^D(r-\|\bx_i-\bx_j\|)\ 
\omega_l(\bx_i,\|\bx_i-\bx_j\|) ,
\end{multline}
with the Dirac distribution $\delta^D(s)$.
On small and intermediate  scales, the global weight $\omega_g$ equals
unity (see Eq.~(\ref{eq:ripley})),  and the Rivolo estimator converges
for $\Delta\rightarrow0$ towards the derivative of the ratio--unbiased
Ripley estimator:
\begin{equation}
4\pi r^2\ \widehat{\rho_1}\ \widehat{g}_3(r) \
\overset{\Delta\rightarrow0}{\longrightarrow} \
\frac{\d}{\d r} \widehat{C}_3(r) .
\end{equation}
Hence, the Rivolo  estimator is approximately  unbiased for radii  $r$
were $\omega_g(r)=1$.

\subsection{The Fiksel and Ohser estimators for $g(r)$}
\label{sect:stoyan-ohser-g}

{}\scite{fiksel:edge}  introduced   the  following  estimator  for the
two--point         correlation        function        (see        also
{}\pcite{ponsborderia:comparing}):
\begin{multline}
\widehat{g_4}(r) = 
\frac{|\CD|}{N^2} \times\\
\times \sum_{i=1}^{N} \sum_{j=1; j\ne i}^{N} 
\frac{\Beins_{[r,r+\Delta]}(\|\bx_i-\bx_j\|)}{4\pi r^2\ \Delta}\ 
\frac{|\CD|}{\gamma_\CD(\bx_i-\bx_j)} ,\\
\text{if } \gamma_\CD(\bx_i-\bx_j) > 0 
\text{ for all } \|\bx_i-\bx_j\|<r . 
\end{multline}
With  arguments presented  in  Sect.~\ref{sect:rivolo}, this estimator
can be derived from the corresponding estimator $\widehat{C_4}(r)$ for
the correlation integral.

Its isotropized  counterpart   $\widehat{g_5}(r)$ is  given  by   (see
{}\pcite{stoyan:fractals} and {}\pcite{ohser:grundlagen}):
\begin{multline}
\widehat{g_5}(r) = \frac{|\CD|^2}{N^2\ 
\overline{\gamma_\CD}(r)} \sum_{i=1}^N\ 
\frac{n_i^\Delta(r)}{4\pi r^2\ \Delta}, 
\text{ for } \overline{\gamma_\CD(r)} > 0 .
\end{multline}
{}\scite{ohser:grundlagen}   use a kernel--based method  instead  of
$n_i^\Delta(r)$ (see Sect.~\ref{sect:remarks}).

\section{Estimators for the two--point correlation 
function $g(r)$ based on $DR$ and $RR$}
\label{sect:pair-estimators}

In  the  cosmological  literature,  estimators for  $g(r)$  are  often
constructed by generating an additional  set of random points.  In the
following  we   consider  $N_{\rm  rd}$   Poisson  distributed  points
$\{\by_j\}_{j=1}^{N_{\rm  rd}}$,   all  inside  the   sample  geometry
$\by_j\in\CD$,   with    the   number   density   $\overline{\rho_{\rm
rd}}=\tfrac{N_{\rm  rd}}{|\CD|}$.   The set  of  the  $N$ data  points
(e.g.~galaxies) is given by $\{\bx_i\}_{i=1}^N$, as before.  We employ
the common notation, and define
\begin{equation}
\label{eq:DD}
DD(r) = \sum_{i=1}^N\ n_i^\Delta(r) ,
\end{equation}
the number of {\em  data--data} pairs with a  distance $[r,r+\Delta]$;
pairs are counted twice.
The number   of  {\em    data--random}   pairs  with a   distance
$[r,r+\Delta]$ is denoted by
\begin{equation}
DR(r) = \sum_{i=1}^N\ dr_i^\Delta(r),
\end{equation}
and
\begin{equation}
dr_i^\Delta(r) = \sum_{j=1}^{N_{\rm rd}}\ 
\Beins_{[r,r+\Delta]}(\|\bx_i-\by_j\|)
\end{equation}
is the number of  {\em random}  points inside  a shell with  thickness
$\Delta$ at a distance $r$ from the {\em data} point $\bx_i$.
Similarly,
\begin{equation}
RR(r) = \sum_{i=1}^{N_{\rm rd}} rr_i^\Delta(r)
\end{equation}
is   the number   of   {\em  random--random} pairs    with a  distance
$[r,r+\Delta]$; pairs are counted twice.   Finally, the number of {\em
random} points inside  a shell with thickness  $\Delta$ at  a distance
$r$ from the {\em random} point $\by_i$ is given by
\begin{equation}
rr_i^\Delta(r) = 
\sum_{j=1, j\ne i}^{N_{\rm rd}}\ \Beins_{[r,r+\Delta]}(\|\by_i-\by_j\|).
\end{equation}

Firstly,     we    show     that    $DR(r)$     and     $RR(r)$    are
Monte--\-Car\-lo--\-ver\-sions    of    well    defined    geometrical
quantities\footnote{These   results  were  independently   derived  by
{}\scite{stoyan:improving}.}.   Secondly,  we  rewrite the  estimators
using  the pair--counts  $DD(r)$, $DR(r)$,  and $RR(r)$,  in  terms of
these  geometric  quantities and  calculate  the  biases entering  the
pair--count estimators.

\subsection{The geometric interpretation of $DR$ and $RR$}

For large $N_{\rm rd}$ and small $\Delta$ we obtain
\begin{equation}
dr_i^\Delta(r) =
\overline{\rho_{\rm rd}}\ \text{area}(\partial\CB_r(\bx_i)\cap\CD)\ \Delta ,
\end{equation}
and therefore
\begin{eqnarray}
\label{eq:DR-1}
DR(r) 
&=& \overline{\rho_{\rm rd}}\ \Delta\
\sum_{i=1}^N \text{area}(\partial\CB_r(\bx_i)\cap\CD) \nonumber \\
&=& 4\pi r^2\Delta\ \overline{\rho_{\rm rd}}\ 
\sum_{i=1}^N \frac{1}{\omega_l(\bx_i,r)} 
\end{eqnarray}
is proportional to the average inverse local weight $\omega_l$ 
(see Eq.~\ref{eq:localweight}).

To  clarify the  geometrical properties  of $RR(r)$  we  re\-write the
set--covariance  (Eq.(\ref{eq:setcov})) as a  Monte--Carlo integration
using  $N_{\rm   rd}$  random  points   $\by_i\in\CD$.   With  $N_{\rm
rd}\rightarrow\infty$ we obtain:
\begin{equation}
\gamma_\CD(\bx) =
\frac{|\CD|}{N_{\rm rd}}\sum_{i=1}^{N_{\rm rd}} \Beins_\CD(\by_i-\bx) .
\end{equation}
After  angular  averaging (see Eq.(\ref{eq:isotrop-setcov})) we insert
an integral over the delta distribution $\delta^D$:
\begin{eqnarray}
\overline{\gamma_\CD}(r) & = &
\frac{|\CD|}{4\pi N_{\rm rd}} \sum_{i=1}^{N_{\rm rd}}\ 
\int_0^{2\pi}\int_0^\pi
\sin(\theta)\d\theta\d\phi \int_0^\infty \d r'\nonumber\\
& & \qquad \times \delta^D(r'-r)\ \Beins_\CD(\by_i+\bx(r',\theta,\phi)) 
\nonumber \\
& = & \frac{|\CD|}{4\pi r^2 N_{\rm rd}} \sum_{i=1}^{N_{\rm rd}} 
\int_{\BR^3} \d^3 z_i 
\nonumber \\
& & \qquad \times \delta^D(\|\bz_i-\by_i\|-r)\ \Beins_\CD(\bz_i) .
\end{eqnarray}
The volume integral in the last line  can be written as a Monte--Carlo
integration:
\begin{equation}
\overline{\gamma_\CD}(r) = \frac{|\CD|^2}{4 \pi r^2 N_{\rm rd}(N_{\rm rd}-1)} 
\sum_{i=1}^{N_{\rm rd}} \sum_{j=1,j\ne i}^{N_{\rm rd}} 
\delta(\|\by_i-\by_j\|-r) .
\end{equation}
For large $N_{\rm rd}$ and small $\Delta$ this results in
\begin{equation}
\overline{\gamma_\CD}(r) = \frac{|\CD|^2}{N_{\rm rd}^2}\ 
\sum_{i=1}^{N_{\rm rd}} \frac{rr_i^\Delta(r)}{4\pi r^2\ \Delta} .
\end{equation} 
Therefore  $RR(r)$  is    proportional to   the   isotropized
set--covariance.
We summarize:
\begin{align}
\label{eq:RR}
RR(r) & = 4\pi r^2 \Delta\ \overline{\rho_{\rm rd}}^2\ 
\overline{\gamma_\CD}(r), \\
\label{eq:DR}
DR(r) & = 4\pi r^2\Delta\ \overline{\rho_{\rm rd}}\ 
\sum_{i=1}^N \frac{1}{\omega_l(\bx_i,r)} .
\end{align}

\subsection{The $DD/RR$ estimator for $g(r)$}

Traditionally, the two--point correlation  function
is estimated by $DD/RR$,
\begin{equation}
\widehat{g_6}(r) = \frac{N_{\rm rd}^2}{N^2} \frac{DD(r)}{RR(r)} .
\end{equation}
From    Eq.~(\ref{eq:DD})     and    (\ref{eq:RR})    we   see,   that
$\widehat{g_6}(r)$  is a Monte--Carlo  version of  the Ohser estimator
$\widehat{g_5}(r)$, which is ratio--unbiased for $\Delta\rightarrow0$.

\subsection{The Davis--Peebles  estimator for $g(r)$}

{}\scite{davis:survey-twopoint} popularized the $DD/DR$ estimator,
\begin{equation}
\widehat{g_7}(r) = \frac{N_{\rm rd}}{N}\ \frac{DD(r)}{DR(r)} .
\end{equation}
{}\scite{landy:bias}  have  shown   that  this  estimator  is  biased.
Rewriting     $\widehat{g_6}(r)$     with    Eq.~(\ref{eq:DD})     and
(\ref{eq:DR-1}) gives
\begin{equation}
\widehat{g_7}(r) = \frac{|\CD|}{N^2}\ 
\frac{\sum_{i=1}^N n_i^\Delta(r)/\Delta}
{\tfrac{1}{N} \sum_{i=1}^N \frac{4\pi r^2}{\omega_l(\bx_i,r)}} .
\end{equation}
A comparison with the Rivolo estimator (Eq.~(\ref{eq:rivolo})), 
\[
\widehat{g_3}(r) = \frac{|\CD|}{N^2} \sum_{i=1}^N 
\frac{n_i^\Delta(r) / \Delta}{\frac{4\pi r^2}{\omega_l(\bx_i,r)}} 
\]
which   is  ratio--unbiased  for    $\Delta\rightarrow0$, reveals  the
geometrical  nature of the    bias.  In $\widehat{g_7}(r)$   the local
weights $\omega_l$ are replaced by  an {\em average} over these  local
weights with the  tacit assumption that the local  weight for a sample
point is  independent of  its relative position  with  respect to  the
boundary, which is unjustified.

\begin{figure}
\begin{center}
\epsfxsize=8cm
\begin{minipage}{\epsfxsize} \epsffile{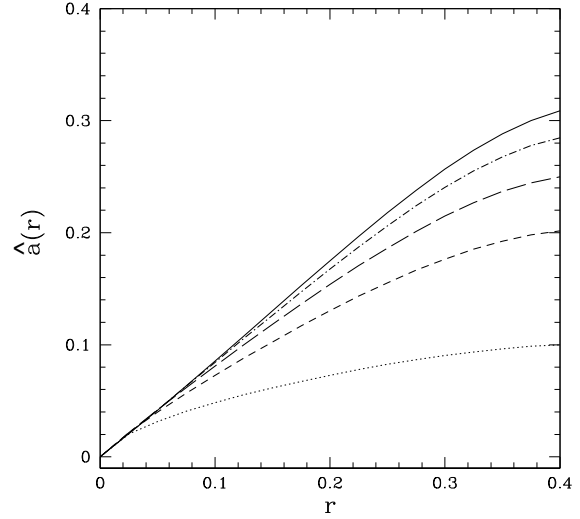} 
\end{minipage} 
\end{center}
\caption{\label{fig:DP-bias} The  average   of $\widehat{a}(r)$   over
10,000 realizations  of a Poisson  process with  $\overline{\rho}=200$
(solid),    a line     segment     process   with  number      density
$\overline{\rho}=200$,  segment  length  $l=0.3$, and  segment density
$\overline{\rho}=1$  (dotted), $\overline{\rho}_s=3$ (short   dashed),
$\overline{\rho}_s=5$ (long dashed), and $\overline{\rho}_s=10$ (short
dashed--dotted).}
\end{figure}

Let us consider the difference
\begin{equation}
\label{eq:DP-rivolo-diff}
\widehat{g_7}(r)-\widehat{g_3}(r) = 
\frac{1}{N}\ \sum_{i=1}^N \frac{n_i^\Delta(r)}{\widehat{\rho_1}\ 
4\pi r^2\Delta}\ A(\bx_i,r), 
\end{equation}
with
\begin{equation}
A(\bx_i,r) = 
\frac{1}{ \tfrac{1}{N} \sum_{j=1}^N \tfrac{1}{\omega_l(\bx_j,r)} } - 
\omega_l(\bx_i,r).
\end{equation}
Fig.~\ref{fig:DP-bias} displays the ensemble average of
\begin{equation}
\widehat{a}(r) = \frac{1}{N} \sum_{i=1}^N A(\bx_i,r)
\end{equation}
and illustrates the bias entering $\widehat{g_7}(r)$.  If we look at a
clustered distribution with  $g(r)\gg1$,  the  bias is  negligible  on
small scales.  However, on large scales, we have $\widehat{a}(r)>0$ of
order    unity.  Since for  a   stationary point   process $g(r)$ also
approaches unity on large  scales,  the bias from $\widehat{a}(r)$  is
important, and $g(r)$ may be overestimated by $\widehat{g_7}(r)$.

Furthermore, $n_i^\Delta(r)$ and $A(\bx_i,r)$ are not independent, and
$\BE\ [\widehat{a}(r)]$ may overestimate  the   true bias, but   since
$n_i^\Delta(r)\ge0$        and                   the              term
$\frac{n_i^\Delta(r)}{\widehat{\rho_1}\     4\pi    r^2\Delta}$   from
Eq.(\ref{eq:DP-rivolo-diff})  is of order unity  on large scales for a
homogeneous point  process,  $n_i^\Delta(r)$ and $A(\bx_i,r)$  have to
conspire,  to  give $\BE\   [\widehat{g_7}(r)-\widehat{g_3}(r)]=0$, if
$\widehat{g_7}(r)$ should be unbiased.

\subsection{The Landy--Szalay estimator for $g(r)$}

\begin{figure}
\begin{center}
\epsfxsize=8cm
\begin{minipage}{\epsfxsize} \epsffile{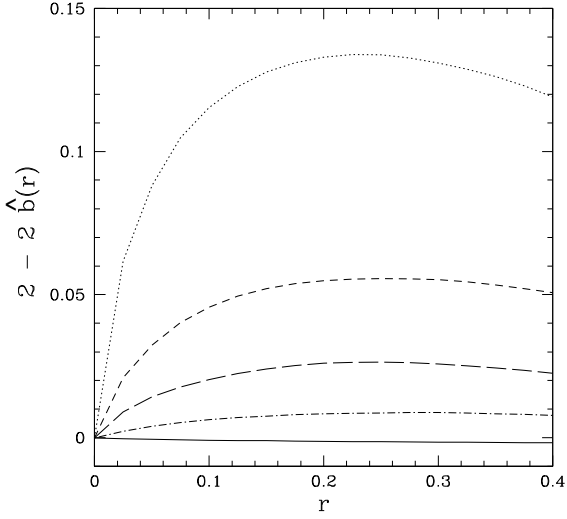} 
\end{minipage} 
\end{center}
\caption{\label{fig:bhat} The average  of the bias $2-2\widehat{b}(r)$
over 10,000  realizations of a Poisson  process in the  unit cube with
$\overline{\rho}=200$  (solid line), and  line segment  processes (see
Sect.~\ref{sect:C-comparison})  with  the  same  number  density,  the
segment  length   $l=0.3$,  and  mean  number  of   lines  per  volume
$\overline{\rho}_s=1$ (dotted),  $\overline{\rho}_s=3$ (short dashed),
$\overline{\rho}_s=5$ (long dashed), and $\overline{\rho}_s=10$ (short
dashed--dotted).}
\end{figure}

{}\scite{landy:bias}  introduced  a new  estimator  for the two--point
correlation function (see also {}\pcite{szapudi:new}):
\begin{equation}
\widehat{g_8}(r) = \frac{N_{\rm rd}^2}{N^2} \frac{DD(r)}{RR(r)} - 
2 \frac{N_{\rm rd}}{N}\frac{DR(r)}{RR(r)} + 2 .
\end{equation}
By   using Eq.~(\ref{eq:RR}) and  (\ref{eq:DR}) and  the definition of
$\widehat{g_6}(r)$ we have 
\begin{equation}
\widehat{g_8}(r) = \widehat{g_6}(r) - 2\ \widehat{b}(r) +2 ,
\end{equation}
with
\begin{equation}
\label{eq:def-lsbias}
\widehat{b}(r) = \frac{N_{\rm rd}}{N}\frac{DR(r)}{RR(r)}
= \frac{\frac{1}{N}\ \sum_{i=1}^N 
\frac{1}{\omega_l(\bx_i,r)} }{\overline{\gamma_\CD}(r)/|\CD| } .
\end{equation}
Since  $\widehat{g_5}(r)$ and   equivalently   $\widehat{g_6}(r)$  are
ratio--unbiased    for   $\Delta\rightarrow0$, $\widehat{g_8}(r)$   is
approximately unbiased only if
\begin{equation}
\BE\ [ \widehat{b}(r)] = 1.
\end{equation}
For a Poisson process in a spherical window this  can be verified from
basic geometric  considerations.    {}\scite{landy:bias}  showed  that
$\widehat{g_8}(r)$   is  ratio--unbiased   for  Poisson   and  binomial
processes in arbitrary windows.
By definition,  neither a Poisson process  nor a binomial process show
large--scale  structures.    To     investigate the   bias    entering
$\widehat{g_8}(r)$ we estimate $\BE\ [\widehat{b}(r)]$ numerically for
the highly structured  line segment process.  In Fig.~\ref{fig:bhat} a
strong bias is visible if only few line segments are inside the sample
geometry.  When   more  and  more   structure  elements  enter,  $\BE\
[\widehat{b}(r)]$ tends towards unity.   Similar to the  properties of
the  $DD/DR$ estimator, this bias   $\widehat{b}(r)$ is unimportant on
small scales  for a clustering process  (with $g(r)\gg1$), as provided
by   the galaxy distribution.  However for   a point distribution with
structures    on   the  size     of      the  sample    (see     e.g.\
{}\pcite{huchra:cfa2s1}), $\widehat{b}(r)$ introduces a bias   towards
higher values in $\widehat{g_8}(r)$ on large scales.

\subsection{The Hamilton estimator for $g(r)$}

{}\scite{hamilton:towards} suggested the following estimator:
\begin{equation}
\widehat{g_9}(r) = \frac{DD(r) RR(r)}{DR(r)^2}.
\end{equation}
With    Eq.~(\ref{eq:DD}),    (\ref{eq:RR}),      (\ref{eq:DR}),   and
(\ref{eq:def-lsbias}) we obtain
\begin{equation}
\widehat{g_9}(r) =  \frac{\widehat{g_7}(r)}{\widehat{b}(r)},
\end{equation}
The Hamilton estimator is unbiased only in the unlikely case where the
biases from $1/\widehat{b}(r)$ and $\widehat{g_7}(r)$ cancel, .

{}\scite{stoyan:improving} found a negative bias in $\widehat{g_9}(r)$
for a Poisson and a Mat\'ern  cluster process.  They attribute this to
an      inappropriate     estimate    of      the     density     (see
Sect.~\ref{sect:improved}).   A   simulation  of  a   Mat\'ern cluster
process, gives a $\BE\ [\widehat{b}(r)]  \approx 1$ as for the Poisson
process,    which suggests  that  mainly the    same  bias  as in  the
Davis--Peebles estimator contributes (see also {}\pcite{landy:bias}).

\section{Improved estimators for $C(r)$ and $g(r)$}
\label{sect:improved}

Recently,    {}\scite{stoyan:improving} proposed several improvements
for  ratio--unbiased estimators of point process statistics. 

With 
\begin{equation}
\kappa(r) = \overline{\rho}\ C(r),
\end{equation}
the density  of point--{\em pairs}  with a distance smaller  than $r$,
ratio--unbiased estimators  of the correlation integral  $C(r)$ may be
written as
\begin{equation}
\label{eq:Cratiounb-definition}
\widehat{C}(r) = \frac{\widehat{\kappa}(r)}{\widehat{\rho}(r)}.
\end{equation}
{\em Unbiased} estimates of $\kappa(r)$ are given by
\begin{align}
\widehat{\kappa_3}(r) = &
\sum_{i=1}^{N}  \sum_{j=1; j\ne i}^{N} 
\Beins_{[0,r]}(\|\bx_i-\bx_j\|)\ \times \\
& \quad \times\ \omega_l(\bx_i,\|\bx_i-\bx_j\|)\ 
\frac{\omega_g(\|\bx_i-\bx_j\|)}{|\CD|} ,\nonumber \\
\widehat{\kappa_4}(r) = &
\sum_{i=1}^{N} \sum_{j=1; j\ne i}^{N} 
\Beins_{[0,r]}(\|\bx_i-\bx_j\|)\ \frac{1}{\gamma_\CD(\bx_i-\bx_j)} \\
\widehat{\kappa_5}(r) = & 
\sum_{i=1}^{N}  \sum_{j=1; j\ne i}^{N} \Beins_{[0,r]}(\|\bx_i-\bx_j\|)\ 
\frac{1}{\overline{\gamma_\CD}(\|\bx_i-\bx_j\|)} .
\end{align}
Using  the {\em unbiased}  estimate $\widehat{\rho_1}=N/|\CD|$  of the
density  $\overline{\rho}$ in  Eq.~(\ref{eq:Cratiounb-definition}), we
recover  the {\em  ratio--unbiased}  estimators $\widehat{C_3}(r)$  to
$\widehat{C_5}(r)$.

{}\scite{stoyan:improving}  showed  that  one   can  do   better.  For
stationary point  processes they   consider  the  following   unbiased
estimate of the density $\overline{\rho}$, also depending on the scale
$r$ under consideration:
\begin{equation}
\label{eq:rho-improved-estimator}
\widehat{\rho_V}(r)  =  
\frac{\sum_{i=1}^N p_V(\bx_i,   r)}{\int_\CD  \d^3x\ p_V(\bx, r)},
\end{equation}
where $p_V(\bx, r)$ is a non--negative weight function. For estimators
of  Ripley's   $K(r)=C(r)/\overline{\rho}$      (see      Appendix~A),
{}\scite{stoyan:improving} employ the volume weight:
\begin{equation}
p_V(\bx,r) = \frac{|\CD\cap\CB_r(\bx)|}{4\pi /3\ r^3 } .
\end{equation}
For     $\mu=3,4,5$  we  define   the   improved {\em ratio--unbiased}
estimators $\widehat{C^i_\mu}(r)$ for the correlation integral
\begin{equation}
\widehat{C^i_\mu}(r) = 
\frac{\widehat{\kappa_\mu}(r)}{\widehat{\rho_V}(r)}.
\end{equation}
A   numerical   comparison,   similar   to  the   one   performed   in
Sect.~\ref{sect:comparisonC},   showed  that   the  variance   of  the
Ripley--\-esti\-mator  $\widehat{C_3}(r)$  is  already  equal  to  its
improved    counterpart    $\widehat{C^i_3}(r)$.    
The improved  estimators $\widehat{C^i_4}(r)$ and $\widehat{C^i_5}(r)$
now show the  same variance as the Ripley--estimator,  hence a smaller
variance   than  the   original   estimators  $\widehat{C_4}(r)$   and
$\widehat{C_5}(r)$.  The  biases do not change between  the normal and
the improved versions of the estimators.

In close analogy  to the estimators  of the correlation  integral, the
estimators    of the  two--point   correlation  function  can also  be
improved.
Consider  the  product density  $\eta(r)  = \rho_2(\bx_1,\bx_2)$  with
$r=\|\bx_1-\bx_2\|$ (see Appendix~A) then
\begin{equation}
\overline{\rho}^2\ g(r) = \eta(r) .
\end{equation}
Ratio--unbiased  estimators  of   the two--point  correlation function
$g(r)$ may be written as
\begin{equation}
\label{eq:gratiounb-definition}
\widehat{g^i_\mu}(r) = 
\frac{\widehat{\eta^\Delta_\mu}(r)}{\widehat{\rho^2}(r)}.
\end{equation}
The estimators  $\widehat{\eta^\Delta_\mu}(r)$ may be defined in terms
of   the   estimators   $\widehat{\kappa_\mu}(r)$  (for  details   see
{}\pcite{stoyan:improving}):
\begin{equation}
\widehat{\eta^\Delta_\mu}(r) = 
\frac{\widehat{\kappa_\mu}(r+\Delta)-\widehat{\kappa_\mu}(r)}{4\pi r^2 \Delta}.
\end{equation}
An  additional complication  enters, since   now  we have to  estimate
$\overline{\rho}^2$,    instead    of    $\overline{\rho}$.    Neither
$(\widehat{\rho})^2$,                                              nor
$\widehat{\rho}(\widehat{\rho}|\CD|-1)/|\CD|$ give  unbiased estimates
of $\overline{\rho}^2$  (the  last  one   is unbiased for   a  Poisson
process). 

Assuming  the  two--point correlation  function  $g(r)$  to be  known,
{}\scite{stoyan:improving}  showed   that  an  unbiased   estimate  of
$\overline{\rho}^2$ is given by
\begin{equation}
\widehat{\rho^2}(r) = \sum_{i=1}^N \sum_{j=1, j\ne i}^N 
\frac{p_V(\bx_i,r)p_V(\bx_j,r)}{g(\|\bx_i-\bx_j\|)}.
\end{equation}
{}\scite{stoyan:improving}   suggest a    self--consistent   iterative
estimation of   both   $g(r)$ and $\overline{\rho}^2$.   
From simulations {}\scite{stoyan:improving}  infer that the estimators
$\widehat{g_4}$ and $\widehat{g_9}$ are biased towards smaller values,
in the case of a Poisson and a Mat\'ern cluster process.  This bias is
reduced in the corresponding improved estimators $\widehat{g^i_4}$ and
$\widehat{g^i_9}$.    In   their   analysis  $\overline{\rho}^2$   was
estimated by  $(\widehat{\rho_S}(r))^2$ where instead  of $p_V(\bx,r)$
the surface weight
\begin{equation}
p_S(\bx,r) = \frac{\text{area}(\CD\cap\partial\CB_r(\bx))}{4\pi\ r^2 } 
\end{equation}
was used in Eq.~(\ref{eq:rho-improved-estimator}).

\section{Correlation integral of IRAS galaxies}
\label{sect:corrint-1.2jy}

We apply the estimators for the correlation to a volume limited sample
with    80\hMpc\   depth, of  the     IRAS  1.2~Jy redshift  catalogue
{}\cite{fisher:irasdata}.  As suggested  by  the  sample  geometry, we
analyse the  northern part (galactic  coordinates) with  412 galaxies,
and the southern part with 376 galaxies, separately.
\begin{figure*}
\begin{center}
\epsfxsize=16cm
\begin{minipage}{\epsfxsize} \epsffile{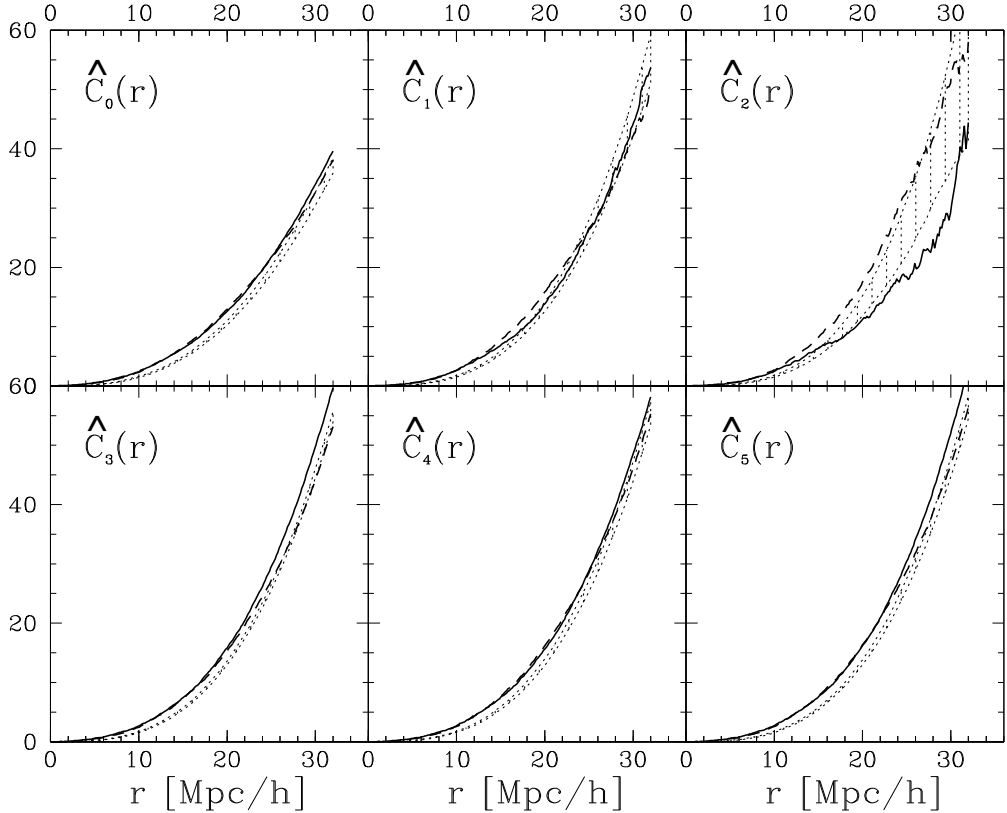} 
\end{minipage} 
\end{center}
\caption{\label{fig:corrjy} Different  estimates of   the  correlation
integral  for a volume   limited  sample of   the IRAS  1.2~Jy  galaxy
catalogue with 80\hMpc\  depth are  shown.   The solid  line marks the
results for the northern, the dashed  line the result for the southern
part, the dotted area marks  the 1$\sigma$ range  of a Poisson process
with the same number density.}
\end{figure*}
In Fig.~\ref{fig:corrjy} we  compare the correlation integral  for the
northern and  southern  parts  with  the  correlation integral  for  a
Poisson process     with  the same  number density,     estimated with
$\widehat{C_0}(r)$ to $\widehat{C_5}(r)$.
As expected, $\widehat{C_0}(r)$   is biased towards lower  values. The
minus--estimators,     especially $\widehat{C_2}(r)$,    show    large
fluctuations on scales above 20\hMpc.
On small  scales out to 10\hMpc\ all  the estimates $\widehat{C_1}(r)$
to $\widehat{C_5}(r)$ of the correlation integral give nearly the same
results  and   are  clearly  above  the   Poisson  result,  indicating
clustering.  Already  {}\scite{lemson:ontheuse} showed, that  on small
scales the  dependence of the  conditional density $\Gamma(r)$  on the
chosen  estimator  is  negligible.   In  Fig.~\ref{fig:corrcompjy}  we
observe a strong scatter in  the estimates of the correlation integral
on  large   scales.   The  minus--estimators   $\widehat{C_1}(r)$  and
$\widehat{C_2}(r)$     deviate      from     $\widehat{C_3}(r)$     to
$\widehat{C_5}(r)$; also the Ripley estimator $\widehat{C_3}(r)$ gives
different    results,    compared    with    $\widehat{C_4}(r)$    and
$\widehat{C_5}(r)$. Whether the  observed correlation integral becomes
consistent with the correlation integral of a Poisson process on large
scales, depends on the chosen estimator.
\begin{figure}
\begin{center}
\epsfxsize=8cm
\begin{minipage}{\epsfxsize} \epsffile{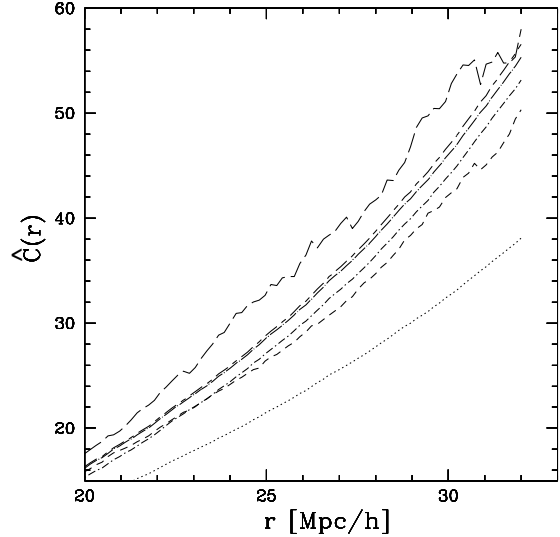} 
\end{minipage} 
\end{center}
\caption{\label{fig:corrcompjy} A    comparison of the   estimators on
large scales   for    the  southern   part   of    the   IRAS  sample:
$\widehat{C_0}(r)$  dotted line; $\widehat{C_1}(r)$ short dashed line;
$\widehat{C_2}(r)$ long   dashed  line;  $\widehat{C_3}(r)$  dotted --
short dashed  line;  $\widehat{C_4}(r)$  dotted --  long  dashed line;
$\widehat{C_5}(r)$ short dashed -- long dashed line.}
\end{figure}

Additionally  to  the differences  between the  estimators, we observe
fluctuations in the correlation integral between  the northern and the
southern       part     (see   also     {}\pcite{martinez:searching}).
{}\scite{kerscher:fluctuations} argued   that  these fluctuations  are
real structural differences between  the northern and southern part of
the sample, observable out to scales of 200\hMpc.

\subsection{A note on scaling}

\begin{figure*}
\begin{center}
\epsfxsize=16cm
\begin{minipage}{\epsfxsize} \epsffile{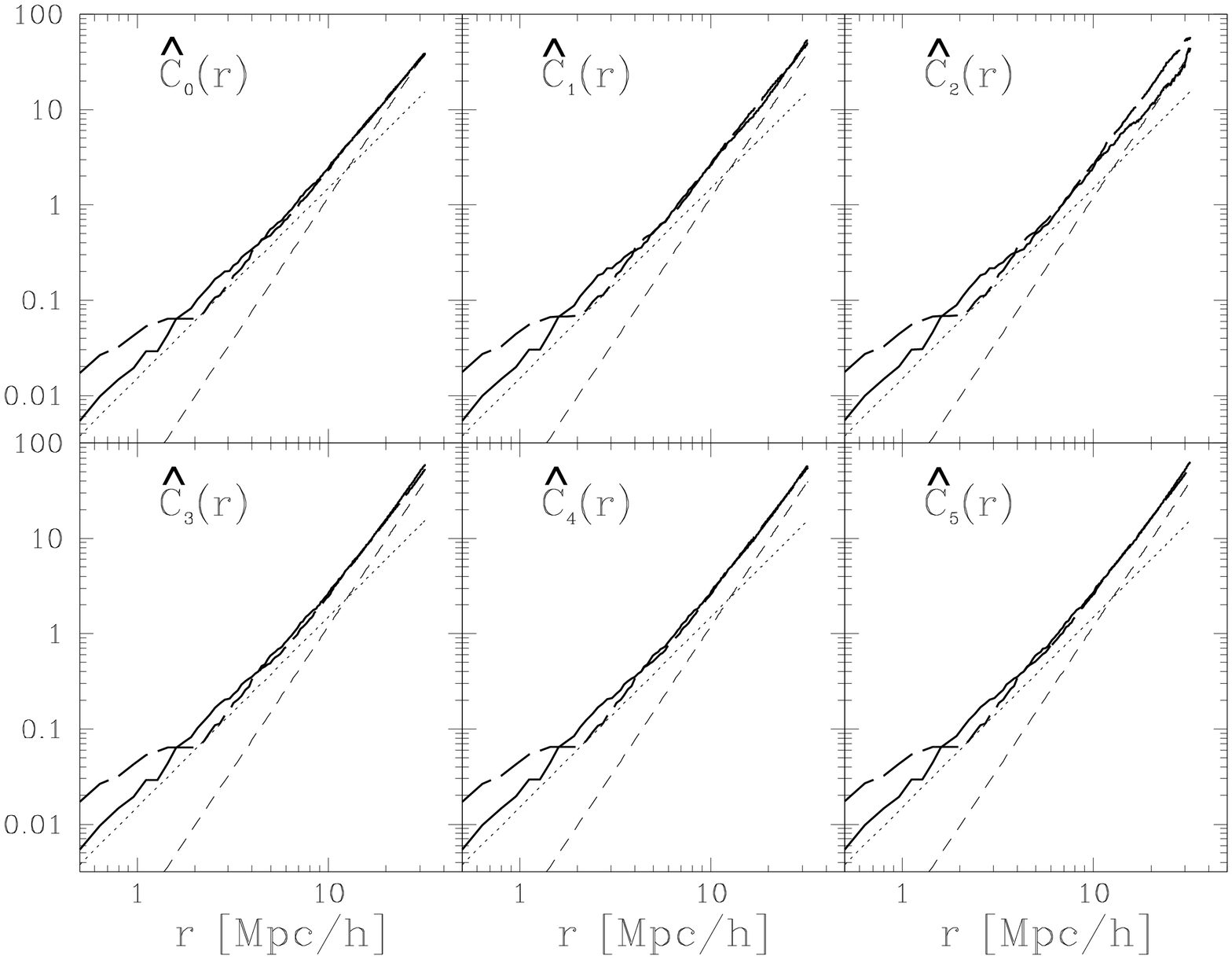} 
\end{minipage} 
\end{center}
\caption{\label{fig:corrlog}  A double  logarithmic plot  of different
estimates of  the correlation integral, for  the northern  part (solid
line)    and the  southern part    (long  dashed line), together  with
functions proportional to $r^2$  (dotted line) and $r^3$ (short dashed
line).}
\end{figure*}

The     Fig.~\ref{fig:corrlog}    displays    the   same   data   from
Fig.~\ref{fig:corrjy} in a double logarithmic plot.
With  all estimators we  observe $C(r)\propto r^D$, $D\approx2$ within
an  approximate scaling regime\footnote{As   we  will argue below,   a
correlation {\em dimension} $D_2$  cannot be reliably extracted from a
scaling regime with roughly one and a  half decades.  Therefore, we do
not   perform a  numerical  fit  to estimate   $D_2$.} up to  at least
10\hMpc.
Above 20\hMpc\  there  seems  to  be  a turnover  towards $D\approx3$.
{}\scite{labini:scale}   argue, that  this turnover    is  due to  the
sparseness of this galaxy catalogue.
In the limit $r\rightarrow0$ the scaling exponent  $D$ is equal to the
correlation dimension $D_2$ {}\cite{grassberger:dimensions}.
Clearly  we  find   approximately  the  same   scaling   properties as
{}\scite{labini:scale}, who analysed this IRAS sample, and a number of
others, using  minus--estimators  equivalent to $\widehat{g_1}(r)$ and
$\widehat{C_1}(r)$.
Similar  results have  been  obtained by {}\scite{martinez:searching},
who   determined  Ripleys      $K(r)=C(r)/\overline{\rho}$  with   the
Ripley--\-estimator, eqivalent to   $\widehat{C_3}(r)$,  for a  volume
limited  sample with 120\hMpc\  depth.  In their Figure~10 the scaling
regime with  $D\approx2$  extends out  to $\approx15$\hMpc, showing  a
turnover towards $D\approx3$ on larger scales.

\begin{figure}
\begin{center}
\epsfxsize=8cm
\begin{minipage}{\epsfxsize} \epsffile{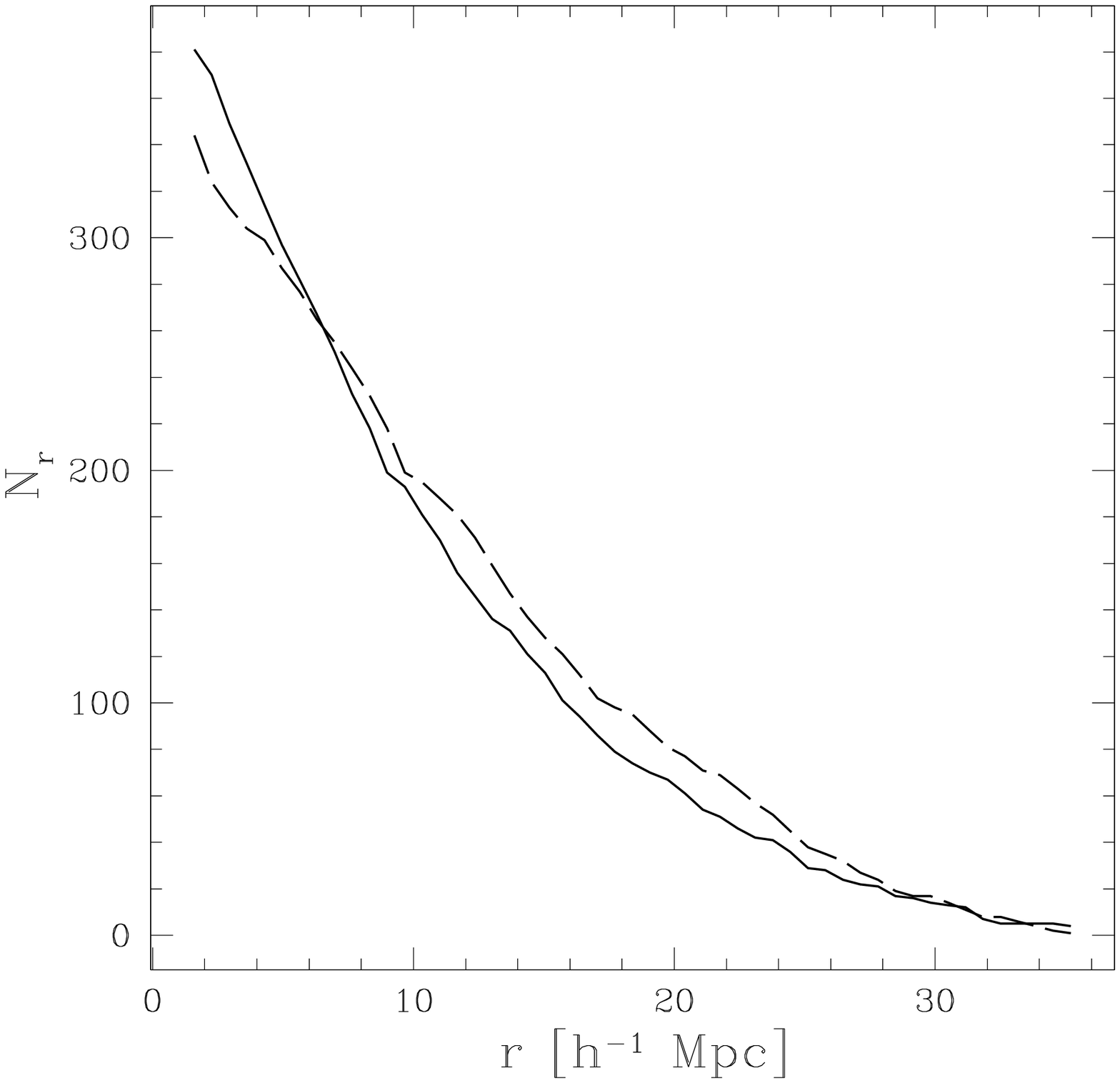} 
\end{minipage} 
\end{center}
\caption{\label{fig:jy-Nr} The   number   $N_r$ of galaxies     with a
distance larger than $r$  to the boundary  of the sample window $\CD$,
for the  northern  part (solid  line) and  the  southern  part (dashed
line).}
\end{figure}
In Fig.~\ref{fig:jy-Nr} we observe that  the number $N_r$ of galaxies,
more  distant than $r$ from  the boundary of   the sample window $\CD$
(see  Eq.~(\ref{eq:Nr-def})),  becomes  critically small,   on  scales
larger  than 20\hMpc.   This leads to   the fluctuations in the  minus
estimators.   Likewise,  the  corrections  from  the  weights in   the
estimators  $\widehat{C_3}(r)$  to $\widehat{C_5}(r)$ become  more and
more  important on scales  larger than  20\hMpc.  Therefore, it is not
clear whether the  trend towards  a  scaling exponent $D\approx3$   on
large scales is  a  true physical one, or  a  result of the  weighting
schemes used.

A lively debate on the extent  of the scaling regime is going on.  See
for      example       the      Princeton      discussion      between
{}\scite{davis:homogeneous}   and   {}\scite{pietronero:fractal},  the
discussions at the Ringberg meeting {}\cite{bender:ringberg}, and more
recently      {}\scite{guzzo:homogeneous},     {}\scite{labini:scale},
{}\scite{mccauley:thegalaxy},     {}\scite{martinez:searching},    and
{}\scite{wu:largescale}.

We  want  to emphasize,  that  two--point measures  are insensitive to
structures on large scales (see the examples in {}\pcite{szalay:walls}
and  {}\pcite{kerscher:regular}).  Therefore, the possible observation
of $C(r)\propto r^3$ on large scales does {\em not} imply, that we are
looking at Poisson distributed points.

\begin{figure}
\begin{center}
\epsfxsize=8cm
\begin{minipage}{\epsfxsize} \epsffile{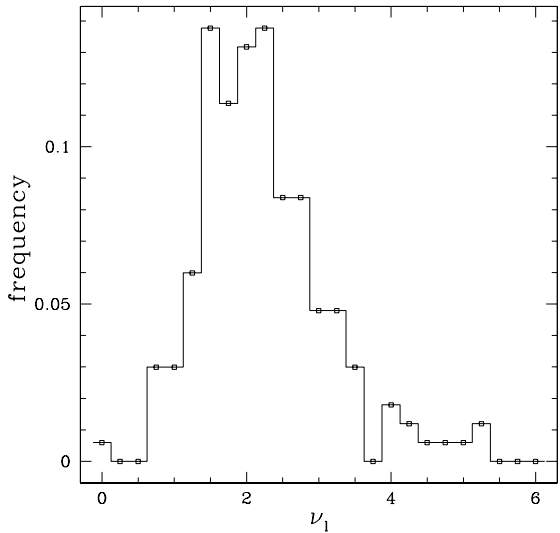} 
\end{minipage} 
\end{center}
\caption{\label{fig:jy-localscaling} The frequency distribution of the
local scaling exponents $\nu_l$.}
\end{figure}

On the other hand, an estimate of the correlation dimension $D_2$ from
one and  a half decades only  is error prone.  To  illustrate this, we
calculate  {\em local} scaling  exponents $\nu_l$  from $N_l(r)\propto
r^{\nu_l}$ (see Eq.~(\ref{eq:local-count}), {}\pcite{stoyan:fractals},
{}\pcite{borgani:scaling} and {}\pcite{mccauley:galaxy}).  We restrict
ourselves to 167 points $\bx_l$ with a distance larger than 12.5\hMpc\
from the  boundary of  the window to  determine $N_l(r)$.   $\nu_l$ is
estimated  using   a  linear  regression   of  $\log(N_l(r))$  against
$\log(r)$.   The frequency  histogram of  the local  scaling exponents
peaks  at $\nu\approx2$,  consistent with  $C(r)\propto r^2$  on small
scales    out    to   10\hMpc,    but    shows    a   large    scatter
(Fig.~\ref{fig:jy-localscaling}).
A constant scaling exponent $D$ may be identified with the correlation
dimension  $D_2$  only,  if  the  scaling regime  of  the  correlation
integral       extends      over       several       decades.       In
{}\scite{grassberger:dimensions}, Fig.~3 a  scaling over 15 decades is
observed,     as    an     unambiguous     trace    of     fractality.
{}\scite{mccauley:galaxy} addresses  the problem of  a limited scaling
regime in the estimation of (multi--) fractal dimensions in detail.

\section{Remarks}
\label{sect:remarks}

\begin{itemize}
\item
The qualitative interpretation of clustering properties is easier with
the  two--point    correlation  function than    with  the correlation
integral.    However, the necessary binning  in  the estimators of the
two--point correlation function may give misleading results, whereas a
quantitative     analysis     with   the   correlation   integral   is
straightforward.
\item
Estimates of  the   two--point  correlation  function  $g(r)$   may be
impaired by shot--noise, due to the finite binning $\Delta$.  However,
no binning is needed in the correlation  {\em integral}, a shot--noise
contribution is visible only at small scales, if at all.
\item 
Sometimes kernel--based methods are used for  the determination of the
two--point correlation function.  The  number  of  points  in a  shell
$n_i^\Delta(r)$ (Eq.~(\ref{eq:def-nidelta})) is replaced by
\[
\sum_{j=1, j\ne i}^N  k_\Delta(r-\|\bx_i-\bx_j\|),
\]
where $k_\Delta$  is a kernel function  of width $\Delta$, satisfying
$k_\Delta(r)=   k_\Delta(-r)\ge0$  and  $\int_{-\infty}^{\infty}\d  r\
k_\Delta(r)=1$     (see       e.g.~\pcite{stoyan:fractals}         and
{}\pcite{ponsborderia:comparing}).
\item
In the Rivolo--  and in the biased Davis--\-Peebles--\-es\-ti\-ma\-tor
we  have set the global  weight $\omega_g$ to  unity, which is correct
for  small  and  intermediate scales.     On  larger scales  we   have
$\omega_g>1$   and   both  estimators   underestimate  the  two--point
correlation function.
\item
On small scales, the weights used in the estimators of the correlation
integral  and  the two--point  correlation  function converge  towards
unity and the biases  $\widehat{a}$ and $\widehat{b}$ converge towards
zero  and  unity,  respectively.   Therefore, all  estimators  of  the
correlation integral and the  two--point correlation function give the
same results on  small scales. However, the quadratic  pole at zero in
the estimators $\widehat{g_3}$ to $\widehat{g_5}$ gives rise to biases
on    very   small    scales   (see    {}\pcite{stoyan:fractals}   and
{}\pcite{ponsborderia:comparing}).
\item 
Only  finite--size   corrections  were   discussed.    Therefore,  the
estimators described are  applicable  to  complete or  volume--limited
samples. Usually a correction for systematic incompleteness effects in
magnitude  limited catalogues  is  performed  by  weighting with   the
inverse   selection   function (see e.g.~{}\pcite{martinez:measures}).
This relies on the {\em assumption}, that the clustering properties of
galaxies are independent of their absolute magnitude.
\item 
Estimators for  the $n$--point  correlation functions, similar  to the
Landy--Szalay--   and    Hamilton--estimators   for   the   two--point
correlation  function  were  introduced by  {}\scite{szapudi:new}  and
{}\scite{jing:threepointlcrs}.   It is not  clear, whether  the biases
found in  the estimators for  the two--point correlation  function are
also present in the  related estimators for the $n$--point correlation
function.   Unbiased estimators  for the  $n$--th moment  measures are
discussed by {}\scite{hanisch:reduction}.
\item
The fit of a straight line to the log--log plot  of the non parametric
estimate of the correlation integral is only one  way to determine the
scaling  properties  of the   point  distribution.  Maximum likelihood
methods are discussed by {}\scite{ogata:maximum}.
\item 
The  attribute {\em (ratio--)  unbiased}  of an  estimator makes sense
only for a stationary point processes.  We emphasize that stationarity
(i.e.\  homogeneity) is a {\em model  assumption}.  It is not possible
to  test global stationarity in  an objective way with one realization
only.  See {}\scite{matheron:estimating} for  a detailed discussion of
the problems inherent in a statistical analysis of {\em one} data set.
\end{itemize}

\section{Conclusions and recommended estimators}
\label{sect:conclusion}

In this  article we are concerned with  the  geometrical nature of the
two--point measures.
As a starting point  we discussed several well--known, ratio--unbiased
estimators of the correlation integral.  From two examples we saw that
all the estimators could reproduce the theoretical  mean values of the
correlation integral  for a Poisson process  and with a small negative
bias  for a  line  segment  process.   The estimators using  weighting
schemes show  a  small   variance,   whereas  the  variance   of   the
minus--estimators becomes prohibitive on large scales.
We investigated the close   relation of the geometrical estimators  of
the two--point correlation function $g(r)$ with  the estimators of the
correlation integral $C(r)$.

Expressing the pair--counts   $DR$ and $RR$  in terms  of  geometrical
quantities  enabled  us   to   calculate   the biases  entering    the
Davis--\-Peebles--\-estimator, the Landy--\-Szalay--\-es\-ti\-ma\-tor,
and the Hamilton--\-estimator.  With simulations of a structured point
process we have quantified these biases:
on small scales they are   unimportant in  the analysis of   clustered
galaxies.  However,  on large  scales the  biases are  not negligible,
especially when only a few structure elements like filaments or sheets
are inside the sample.

As a real--life example we applied the estimators  to a volume limited
sample   extracted   from   the    IRAS  1.2~Jy   galaxy     catalogue
{}\cite{fisher:irasdata} with 80\hMpc\ depth.
On  scales  up  to   10\hMpc\  the  estimators  $\widehat{C_1}(r)$  to
$\widehat{C_5}(r)$  gave nearly  identical results,  and the  shape of
$C(r)$ is well determined in  the northern and southern part (galactic
coordinates) of the sample separately.  However, on scales larger than
20\hMpc\ the  results differ,  not only between  the minus--estimators
and  the estimators  using weighting  schemes, but  also  between {\em
ratio--unbiased} estimators using different weighting schemes.
In a scaling analysis we found a $C(r)\propto r^D$ with $D\approx2$ up
to 10\hMpc,  and a possible  turnover to $D\approx3$ on  scales larger
than 20\hMpc.   However, the extent  of this scaling regime  cannot be
reliably determined from this galaxy sample.  Since the scaling regime
is roughly one and a half decades only, an estimate of the correlation
dimension  $D_2$ from  the scaling  exponent $D$  is  unreliable.  The
large  scatter  seen in  the  local  scaling  exponents reflects  this
uncertainty.
We    could     also    confirm    the     fluctuations    found    by
{}\scite{kerscher:fluctuations}  in the clustering  properties between
the   northern  and   southern   parts  of   the   sample  (see   also
{}\pcite{martinez:searching}).

These  large scale fluctuations and  the differences  in the estimated
correlation integral  suggest that  we  have to   wait  for  the  next
generation surveys, like the SDSS and the 2dF, if we want to determine
the two--point measures of galaxies unambiguously on large scales.

\subsection{Recommended estimators}

A general recommendation is that  one should compare the results of at
least two  estimators. Since estimators of  the two--point correlation
function are  ratio--unbiased only in the impracticable  limit of zero
bin width,  statistical tests should  be based on  integral quantities
like the correlation integral $C(r)$ or the $L(r)$ function defined in
Appendix~A.

On small scales  the weights in the estimators  converge towards unity
and  also  the  biases  become  negligible for  the  clustered  galaxy
distribution. This  is confirmed  by our analysis  of the  IRAS 1.2~Jy
catalogue, where all estimators of $C(r)$ give nearly the same results
on scales smaller than 10\hMpc.
Therefore,  on small and  intermediate scales,  i.e.\ on  scales where
$\xi_2(r)$  is of  the  order or  larger  than unity,  the {\em  best}
estimator is the one with the smallest variance and the smallest bias.
For   the  correlation   integral   this  is   the  Ripley   estimator
$\widehat{C_3}$.   For  complicated  sample geometries  the  numerical
implementation of $\widehat{C_4}$ is simpler than $\widehat{C_3}$ (see
Appendix~B).   Additionally, the variance  of $\widehat{C_4}$  is only
slightly increased, and the assumption of isotropy is not entering the
construction of this estimator.

On large scales, i.e.\ on  scales with $\xi_2(r)<1$, the comparison of
the  results obtained  with different  ratio--unbiased  estimators may
serve as  an internal consistency check, and  provide conclusive means
to judge the reliability of the estimates.
The  scale at  which significant  differences  between ratio--unbiased
estimators are found can be used  to define a scale of reliability for
the sample under consideration.
In   particular,   a   comparison  between   the   Ripley--\-estimator
$\widehat{C_3}$  and the  Ohser--\-Stoyan--\-estimator $\widehat{C_4}$
is  useful,  since the  assumption  of  isotropy  does not  enter  the
construction of the Ohser--Stoyan estimator.
A  final  comparison  with  the minus--estimator  $\widehat{C_1}$  can
illustrate the reliability of the results on large scales.

Guided by our  analysis of estimators for the  correlation integral we
expect  that the  estimators  of the  two--point correlation  function
behave similar and that one may use either the Rivolo $\widehat{g_3}$,
or  the Fiksel  estimator  $\widehat{g_4}$ on  small and  intermediate
scales.   On very  small  scales the  quadratic  pole at  zero in  the
estimators  $\widehat{g_3}$  to  $\widehat{g_5}$  can lead  to  biases
{}\cite{stoyan:fractals}.  No estimator of the correlation integral is
impaired by this pole.
Similarly   to  the   correlation   integral  a   comparison  of   the
Rivolo--estimator      $\widehat{g_3}$,      the     Fiksel--estimator
$\widehat{g_4}$, and the minus--estimator $\widehat{g_1}$ may serve as
an internal consistency check on large scales.

In Sect.~\ref{sect:pair-estimators} we  discussed how biases enter the
Davis--Peebles estimator  $\widehat{g_7}$, the Landy--Szalay estimator
$\widehat{g_8}$,  and  the   Hamilton  estimator  $\widehat{g_9}$  for
arbitrary  stationary and  isotropic point  processes.   We quantified
them for  a line  segment process.  The  relevance of these  biases in
cosmological  situations, and  a comparison  of the  variances  of the
pair--count   estimators  with  the   variances  of   the  geometrical
estimators will be subject of future work.

\section*{Acknowledgements}

It is  a pleasure to   thank Claus Beisbart,  Thomas Buchert, Dietrich
Stoyan, Roberto Trasarti--\-Battistoni,  Jens Schmal\-zing and Herbert
Wagner for useful discussions and comments.
My  interest   in this   topic  emerged from  discussions with  Vicent
J. Mart\ii  nez, Joe McCauley, Mar\ii  a Jes\'us Pons--Border\ii a and
Sylos Labini.
Special  thanks to  the referee  Istvan Szapudi  for his  comments.
This work  was supported by  the Sonderforschungsbereich SFB 375 f\"ur
Astroteilchenphysik der Deutschen Forschungsgemeinschaft.


\providecommand{\bysame}{\leavevmode\hbox to3em{\hrulefill}\thinspace}

\appendix

\section{Two--point measures}
\label{sect:twopoint-measures}

In  this appendix    we    summarize common two--point measures.

The product density\footnote{In the statistical literature the product
density  $\rho_2(\bx_1,\bx_2)$ is defined as   the Lebesgue density of
the  second factorial moment measure (e.g.~\pcite{stoyan:stochgeom}).}
\begin{equation}
\rho_2(\bx_1,\bx_2)\d V(\bx_1)   \d V(\bx_2)
\end{equation}
is the   probability to
find a point in  the infinitesimal volume $\d  V(\bx_1)$ {\em and}  in
$\d V(\bx_2)$.  
In the following  we assume that the point  process is stationary  and
isotropic, hence the  statistical properties  of ensemble averages  do
not depend on the specific location and orientation in space.  In this
case   $\rho_2(\bx_1,\bx_2)$   only     depends   on  the     distance
$r=\|\bx_1-\bx_2\|$ of the two points:
\begin{equation}
\overline{\rho}^2 g(r) = \overline{\rho}^2 (1+\xi_2(r)) 
= \rho_2(\bx_1,\bx_2).
\end{equation}
The correlation integral $C(r)=\int_0^r\d s\ \overline{\rho}\ 4\pi s^2
g(s)$ is related to Ripley's  $K$--function (also known as the reduced
second moment measure {}\cite{stoyan:stochgeom}) by
\begin{equation}
C(r)=\overline{\rho}K(r).
\end{equation}
For statistical test, often the $L(r)$ function is used:
\begin{equation}
L(r)= \left( \frac{K(r)}{r^3 4\pi/3} \right)^{1/3}.
\end{equation}
Care has to be  taken, since the  definition  of $L(r)$ is not  unique
throughout the literature.
In some applications the integrated normed cumulant is considered:
\begin{equation}
J_3(r)=\int_0^r\d s\ s^2\ \xi_2(s).
\end{equation}
Clearly, $C(r)=\overline{\rho}\frac{4\pi}{3}\ r^3 +   \overline{\rho}\
4\pi\ J_3(r)$.
{}\scite{coleman:fractal}  use 
\begin{equation}
\Gamma^\star(r)=\frac{C(r)}{r^3 4\pi/3}  \text{  and  }
\Gamma(r)=\overline{\rho}\ g(r).
\end{equation}

Another common tool is the variance of cell counts.  We are interested
in the  fluctuations  in the  number  of points $N(\CC)$  in a spatial
domain   $\CC$.   The variance   of    $N(\CC)$  is  given  by    (see
e.g.~{}\pcite{stoyan:fractals})
\begin{eqnarray}
\BV[N(\CC)]
&=& \BE\ [N(\CC)^2] - \BE\ [N(\CC)]^2 = \\[1ex]
&=& \int_\CC\d^d x_1 \int_\CC\d^d x_2\ \rho_2(\bx_1,\bx_2)  + 
\overline{\rho}|\CC| - (\overline{\rho}|\CC|)^2 . \nonumber 
\end{eqnarray}
$\BE$ is the ensemble average, i.e.\ the average over different 
realizations. For a Poisson process $\BV[N(\CC)]=\overline{\rho}|\CC|$. 
Also $\sigma(r)^2$, the  fluctuations  in excess  of Poisson inside  a
sphere $\CB_r$ with radius $r$ are considered:
\begin{equation}
\BV[N(\CB_r)] = 
\overline{\rho}|\CB_r| +\sigma(r)^2 (\overline{\rho} |\CB_r|)^2 .
\end{equation}
Hence,
\begin{align}
\sigma(r)^2 & =
\frac{1}{|\CB_r|^2}\int_{\CB_r}\int_{\CB_r}\d^3 x\d^3y\ 
\xi_2(\|\bx-\by\|) \nonumber \\
& = \frac{C(r)}{\overline{\rho} |\CB_r|} -1 = (L(r))^3 -1 .
\end{align}

Often spectral methods are used. The power  spectrum can be defined as
the Fourier transform of the normed cumulant $\xi_2$:
\begin{equation}
P(\bk) = \frac{1}{(2\pi)^3}\ \int_{\BR^3} \d^3x\ 
\e^{-i\bk\cdot\bx}\ \xi_2(\|\bx\|)  .
\end{equation}
{}\scite{newman:redshift}  discuss problems  in the  estimation of the
power--spectrum.

\section{Implementation}
\label{sect:implementation}

In this Appendix we give  a short description of the implementation of
the estimators.

\subsection{Minus estimators:}
The main  computational problem  is to determine  the distance  from a
galaxy to  the boundary  of the sample,  or equivalently,  whether the
galaxy is inside the shrunken  window $\CD_{-r}$. No general recipe is
available  and  the  implementation  depends on  the  specific  survey
geometry under consideration.

\subsection{Ripley and Rivolo estimators}
For  both the  Ripley and  the Rivolo  estimators  $\widehat{C_3}$ and
$\widehat{g_3}$    we   have   to    calculate   the    local   weight
Eq.~(\ref{eq:localweight}):
\[
\omega_l(\bx_i,s) =
\begin{cases}
\frac{4\pi s^2}{\text{area}(\partial\CB_{s}(\bx_i)\cap\CD)} 
& \text{for } \partial\CB_{s}(\bx_i)\cap\CD \ne \emptyset, \\
0 & \text{for } \partial\CB_{s}(\bx_i)\cap\CD = \emptyset,
\end{cases}
\]
is inversely proportional to the part  of the surface of a sphere with
radius $s$ drawn  around the point $\bx_1$ which  is inside the survey
geometry $\CD$ (see Fig~\ref{fig:local-weight}).
For   a  cuboid   sample   {}\scite{baddeley:3dpoint}  give   explicit
expressions.    In  our   calculations  we   discretized   the  sphere
$\CB_s(\bx_i)$,  and approximated  $\omega_l(\bx_i,s)$ by  the inverse
fraction   of   surface    elements   inside   the   sample   geometry
$\CD$. Equivalently, a random distribution of points on the sphere may
be used.  In both approaches, we  need a fast method to determine if a
point is inside the sample $\CD$.
{}\scite{rivolo:two-point}  suggested to  count the  number  of random
points  inside $\CD$ in  the shell  of radius  $s$ and  width $\Delta$
around $\bx_i$ to estimate
\[
\overline{\rho_{\rm   rd}}\  \text{area}(\partial\CB_s(\bx_i)\cap\CD)\
\Delta .
\]

Explicit expressions for the global weight 
\[
\omega_g(s) =
\frac{|\CD|}{|\{\bx\in\CD~|~\partial\CB_s(\bx)\cap\CD\ne\emptyset\}|} ,
\]
for a  cuboid sample can be found  in {}\scite{baddeley:3dpoint}.  For
more general  sample geometries it seems necessary  to use Monte-Carlo
methods. In  our calculations we  considered only radii $s$  for which
$\omega_g(s)=1$ is fulfilled.

\subsection{Ohser--Stoyan type estimators}
For  the  Ohser--Stoyan   $\widehat{C_4}$  and  the  Fiksel  estimator
$\widehat{g_4}$,   we   have    to   calculate   the   set--covariance
Eq.~(\ref{eq:setcov}):
\[
\gamma_\CD(\bx) = |\CD \cap \CD+\bx|.
\]
We  obtain  for  a  vector  $\bx=(x,y,z)$ and  a  cuboid  sample  with
side lengths $L_x>|x|, L_y>|y|, L_z>|z|$
\begin{equation}
\gamma_\CD(\bx) = (L_x-|x|)(L_y-|y|)(L_z-|z|),
\end{equation}
and for a spherical sample with Radius $R$ and $r=\|\bx\|<R$
\begin{equation}
\gamma_\CD(\bx) = \frac{4\pi}{3}
\left(R^3 -\frac{3}{4}rR^2 + \frac{1}{16} r^3\right).
\end{equation}
For more general  sample geometries one has to  rely on a Monte--Carlo
method:  draw   random  points  $\by_i$  inside   $\CD$  and  estimate
$\gamma_\CD(\bx)/|\CD|$ by  the fraction of  points $\by_i+\bx$ inside
the sample $\CD$.

The isotropized set--covariance $\overline{\gamma_\CD}(r)$ used in the
estimators $\widehat{C_5}$ and  $\widehat{g_5}$ can be calculated from
$\gamma_\CD(\bx)$.  For a cuboid sample we obtain
\begin{multline}
\overline{\gamma_\CD}(r) = 
L_x L_y L_z -\frac{r}{2}(L_x L_y+L_x L_z+L_y L_z) + \\
+\frac{2r^2}{3\pi}(L_x+L_y+L_z) - \frac{r^3}{4\pi},
\end{multline}
and            for       a                spherical             sample
$\overline{\gamma_\CD}(\|\bx\|)=\gamma_\CD(\bx)$. 
Again,  for  more general  sample  geometries one  has  to  rely on  a
Monte--Carlo  method:  consider  randomly distributed  points  $\by_i$
inside  $\CD$ and  unit vectors  $\bu_j$ randomly  distributed  on the
sphere. Now estimate  $\overline{\gamma_\CD}(r)/|\CD|$ by the fraction
of points $\by_i+r\bu_j$ inside the sample $\CD$.
Another possibility is to use the number of random point pairs $RR(r)$
inside the  sample to  estimate $4\pi r^2  \Delta\ \overline{\rho_{\rm
rd}}^2\ \overline{\gamma_\CD}(r)$ according to Eq.~(\ref{eq:RR}).

\subsection{$DD$, $DR(r)$, and $RR$:}
$DD(r)$,  $DR(r)$,   and  $RR(r)$   are  the  number   of  data--data,
data--random,  and  random--random point  pairs  inside  $\CD$ with  a
distance  in $[r,r+\Delta]$. Point  pairs in  $DD(r)$ and  $RR(r)$ are
counted twice.
Care has  to be  taken to  use enough random  points $N_{\rm  rd}$:

For  small  $r$  the  value  of  the  iso\-tro\-pized  set--covariance
$\overline{\gamma_\CD}(r)$ is close to  $|\CD|$, and $RR(r)$ has to be
approximately    $4\pi   r^2    \Delta   N_{\rm    rd}^2/|\CD|$   (see
Eq.~(\ref{eq:RR})).   If significant deviations  for small  $r$ occur,
the number of random points $N_{\rm rd}$ should be increased.

At small scales the  local weight $\omega_l(\bx_i,r)$ equals unity for
most of  the points $\bx_i$. From Eq.~(\ref{eq:DR})  we recognize that
$DR(r)$ is approximately $4\pi  r^2 \Delta N_{\rm rd} N/|\CD|$.  Again
a significant deviation indicates that more random points are needed.

\end{document}